\documentclass[journal,twoside]{IEEEtran}
\usepackage{stmaryrd}
\usepackage{amsfonts}
\usepackage{amssymb}
\usepackage{stfloats}
\usepackage{hyperref}
\usepackage[square, comma, sort&compress, numbers]{natbib}
\usepackage{graphicx}
\usepackage{psfrag}
\usepackage{subfigure}
\usepackage{amsmath,amsthm}
\usepackage{array}
\usepackage{eurosym}
\usepackage{setspace}
\usepackage{float}
\usepackage{multirow}
\usepackage{booktabs}
\usepackage{algorithm}
\usepackage{algorithmic}
\usepackage{threeparttable}
\usepackage{color}
\usepackage{booktabs}
\usepackage{makecell}
\usepackage{threeparttable}
\usepackage{colortbl}
\renewcommand{\algorithmicrequire}{\textbf{Input:}}
\renewcommand{\algorithmicensure}{\textbf{Output:}}
\definecolor{mygray}{gray}{.9}

\begin{document}
\title{On the Low-Complexity, Hardware-Friendly Tridiagonal Matrix Inversion for Correlated Massive MIMO Systems}
\newtheorem{Thm}{Theorem}
\newtheorem{Lem}{Lemma}
\newtheorem{Cor}{Corollary}
\newtheorem{Def}{Definition}
\newtheorem{Exam}{Example}
\newtheorem{Alg}{Algorithm}
\newtheorem{Prob}{Problem}
\newtheorem{Rem}{Remark}
\renewcommand\thesubsection{\thesection.\Alph{subsection}}
\renewcommand{\multirowsetup}{\centering}
\renewcommand{\algorithmicrequire}{\textbf{Input:}}
\renewcommand{\algorithmicensure}{\textbf{Output:}}

\author{Chuan~Zhang,~\IEEEmembership{Member,~IEEE},
        Xiao~Liang,~
        Zhizhen~Wu,~
        Feng~Wang,~
        Shunqing~Zhang,~\IEEEmembership{Senior Member,~IEEE},
        Zaichen~Zhang,~\IEEEmembership{Senior Member,~IEEE},
        and~Xiaohu~You,~\IEEEmembership{Fellow,~IEEE}
\thanks{Chuan Zhang, Xiao Liang, Zhizhen Wu, Feng Wang, Shunqing Zhang, and Xiaohu You are with the National Mobile Communications Research Laboratory, Southeast University, Nanjing, China. Email: \{chzhang, xiao\_liang, jasonwu, wfeng, zczhang, xhyu\}@seu.edu.cn.
Shunqing Zhang is with Shanghai Institute for Advanced Communications and Data Science, Shanghai University, Shanghai, China. Email: shunqing@shu.edu.cn. Chuan Zhang and Xiao Liang contributed equally to this work. \emph{(Corresponding author: Chuan Zhang.)}
}
}

\markboth{IEEE Transactions on xxx, ~xxxx}
{X. Liang \MakeLowercase{\textit{et al.}}: On Low-Complexity, Hardware-Friendly Tridiagonal Matrix Inversion for Correlated Massive MIMO Systems}

\maketitle

\begin{abstract}
In massive MIMO (M-MIMO) systems, one of the key challenges in the implementation is the large-scale matrix inversion operation, as widely used in channel estimation, equalization, detection, and decoding procedures. Traditionally, to handle this complexity issue, several low-complexity matrix inversion approximation methods have been proposed, including the classic Cholesky decomposition and the Neumann series expansion (NSE). However, the conventional approaches failed to exploit neither the special structure of channel matrices nor the critical issues in the hardware implementation, which results in poorer throughput performance and longer processing delay. In this paper, by targeting at the correlated M-MIMO systems, we propose a modified NSE based on tridiagonal matrix inversion approximation (TMA) to accommodate the complexity as well as the performance issue in the conventional hardware implementation, and analyze the corresponding approximation errors. Meanwhile, we investigate the VLSI implementation for the proposed detection algorithm based on a Xilinx Virtex-7 XC7VX690T FPGA platform. It is shown that for correlated massive MIMO systems, it can achieve near-MMSE performance and $630$ Mb/s throughput. Compared with other benchmark systems, the proposed pipelined TMA detector can get high throughput-to-hardware ratio. Finally, we also propose a fast iteration structure for further research.
\end{abstract}


\section{Introduction}\label{sec:intro}
\IEEEPARstart{M}{assive} MIMO (or large-scale MIMO, M-MIMO), as an emerging technology employing hundreds of antennas at the base station (BS) to serve a relatively small number of users simultaneously, is a promising technology for the next generation wireless communications \cite{Marzetta:12,ChengXiang:14}. By adding multiple antennas, degrees of freedom in wireless channel can be offered to accommodate more information data. Hence, a significant performance improvement can be obtained in terms of data rate, reliability, spectral efficiency, and energy efficiency \cite{Rusek:13}.

Unfortunately, the advantages of M-MIMO systems come at the cost of the significantly increased computational complexity at the BS. One of the most challenging issues is the low-complexity signal detection algorithm in both uplink and downlink directions \cite{Rusek:13}. For the optimal signal detection algorithm such as the maximum-likelihood (ML) detection \cite{ML:verdu1986minimum}, its complexity increases exponentially with the number of terminal antennas \cite{Exp:van1981another}, which entails prohibitive complexity for the M-MIMO detection. The fixed-complexity sphere decoding (SD) \cite{Burg:05,Barbero:08} would also incur unaffordable complexity with the scaling up dimension or the high modulation order. In \cite{Rusek:13}, low-complexity sub-optimal linear detection schemes such as zero-forcing (ZF) \cite{ZF:van1975} and minimum mean square error (MMSE) \cite{MMSE:shnidman1967} have been investigated. Compared to other more sophisticated methods such as belief propagation \cite{BP:hu2008,BP:som2010} and iterative solvers \cite{Iter:hochwald2003,Iter:moher1998,Iter:sugiura2012}, the linear detection methods are favorable especially for practical system designers not only for their low complexity and implementation convenience, but also for their close-to-optimal performance \cite{Rusek:13}. When the number of antennas increases without bound, thermal noise, intra-cell interference, fast fading, and channel estimation errors \cite{Ngo:13}. Moreover, if the number of BS antennas ($N$) is much larger than that of user antennas ($K$), i.e., $N \gg K$, the simplest linear detectors are one of the best choices with respect to performance/complexity tradeoff \cite{Rusek:13}.

Therefore, recent papers \cite{MichaelWu:14,Prabhu:13,lund2014iscas,Kammoun:17,MichaelWu:13} studied linear detection and linear precoding under the optimistic assumption about the propagation conditions and the number of antennas. Some prototype M-MIMO systems also adopt linear algorithms considering the performance and implementation convenience \cite{shepard2013argosv2,nikaein2014openairinterface,choubey2016introducing,lund2017mimo,jin2017chinacom}. All those linear algorithms have to address the key issue of an unfavorable filtering matrix inversion, for its increasing computational complexity in the M-MIMO systems. With the increased size of the channel gain matrix $\mathbf{H}$, the property of channel hardening, that the concern matrix $\mathbf{H}^{H}\mathbf{H}$ tends to be diagonal dominant \cite{Rusek:13,Matthaiou:10}, has inspired recent research groups \cite{MichaelWu:14,lund2014iscas} to propose the low-complexity matrix inverse approximation (MIA) methods.

Most of the studies \cite{MichaelWu:14,Prabhu:13,lund2014iscas,Kammoun:17,MichaelWu:13} referred to above assume that the channels are independent and identically distributed (i.i.d.) Gaussian MIMO channels. However, in practice, the channel vectors for different users are generally correlated because the antennas are not sufficiently separated or the propagation environment does not offer enough scattering \cite{Ngo:13,Hoydis:13}. Spatial correlation between antennas is an important factor that affects the design and performance of MIMO systems, which specially matters for M-MIMO systems, owing to the number of antennas scaling up to an unprecedented magnitude \cite{Rusek:13}. For the scenarios of less favorable propagation environment where the channel is of high correlation or the ratio $\beta$ ($=N/K$) is low, the study in the filtering matrix inversion and its efficient hardware implementation have not been considered based on the best knowledge of the authors.

In this work, we address the complexity issue of the filtering matrix inversion associated with linear detection in the M-MIMO systems. We consider a single-cell multi-user system model, which accounts for antenna correlation, ratio between BS antennas and users, and path loss in the uplink. We analyze the residual error induced by the matrix inversion approximation (MIA) method relying on NSE under the less favorable propagation environment. We propose an iterative MIA method also based on NSE, which can converge to an arbitrary $L$-terms approximation without extra hardware cost compared to the recent works. Considering the residual error and the computational complexity, we analyze the initial matrix $\mathbf{X}$ and then develop an optimizational tridiagonal matrix approximation (TMA) method to improve the convergence performance. We present corresponding VLSI architecture and show the reference FPGA results, which enables us to show the hardware efficiency compared to conventional methods. Finally, we provide the solution considering the performance/complexity tradeoff via presented designs.

\textbf{Notation}:
The operations, $(\cdot)^*$, $(\cdot)^T$, $(\cdot)^H$, $\mathrm{Tr}(\cdot)$, $|\cdot|$, and $\mathbb{E}\{\cdot\}$, denote the conjugate, transpose, conjugate transpose, trace, absolute operator, and expectation, respectively. For a matrix $\mathbf{A}$, the entry in the $i^{\text{th}}$ row and $j^{\text{th}}$ column of $\mathbf{A}$ is denoted by $a_{ij}$. $\|\mathbf{A}\|_{F}$ denotes the Frobenius norm of matrix $\mathbf{A}$. For a vector $\mathbf{a}$, the entry in the $k^{\text{th}}$ is denoted by $a_{k}$; $\|\mathbf{a}\|_{p}$ denotes the $l^{p}$-norm of vector $\mathbf{a}$, such that
\begin{equation}
  \left\|\mathbf {a} \right\|_{p}:={\bigg (}\sum _{i=1}^{n}\left|a_{i}\right|^{p}{\bigg )}^{1/p}.
\end{equation}
$\mathbf{I}_{K}$, $\mathbf{1}_{K}$, and $\mathbf{0}_{K}$ refer to the $K\times K$ identity, full-ones, and zero matrices, respectively. \textcolor{black}{A matrix consists of the entries $a_{ij}, \forall (i-j)=k$ of $\mathbf{A}$ is denoted as $\textrm{diag}_{k}(\mathbf{A})$\footnote{Then $\mathbf{A}=\sum_{k=-(K-1)}^{K-1}{\textrm{diag}_{k}(\mathbf{A})}$.}.}

\section{Preliminaries}\label{sec:sce_tar}
\subsection{System Model}
Consider an uplink M-MIMO system with a single base station (BS) and $K$ user terminals (UTs), where the BS is equipped with $N$ antennas and each UT is equipped with single antenna only\footnote{UTs with multiple antennas can be decomposed into several single antenna cases or can be generalized from single-antenna UT case directly \cite{jafar2004transmitter,rhee2003capacity,kotecha2003capacity}.}. Denote $s_k$ to be the transmitted signal of the $k^{\text{th}}$ UT with a normalized expected total transmit power $\mathbb{E}\{|s_{k}|^{2}\}=1$. \textcolor{black}{The received signal at the BS side, $\mathbf{y} \in \mathbb{C}^{N \times 1} $, can be expressed as}
\begin{equation}\label{eqn:rec_sig}
\mathbf{y} = \sum\nolimits_{k=1}^{K} \mathbf{H}_k s_k + \mathbf{n},
\end{equation}
where $\mathbf{H}_k \in \mathbb{C}^{N \times 1}$ denotes the channel condition between the BS and the $k^{\text{th}}$ UT. $\mathbf{n}$ is the additive white Gaussian noise (AWGN) with zero mean and normalized variance $\eta^2$.

In the conventional uplink M-MIMO system analysis \cite{Marzetta:12}, only the geometric attenuation is adopted, whereas the antenna correlation issue for massive receiving antennas is ignored. In this paper, we apply the Kronecker model as proposed in \cite{kermoal:2002} and obtain
\begin{equation}\label{eqn:channel}
\mathbf{H}_k = \sigma_k^{1/2} \mathbf{R}^{1/2} \widetilde{\mathbf{H}}_k,
\end{equation}
where $\sigma_k^{1/2}$ is the path-loss component between the BS and the $k^{\text{th}}$ UT, and $\mathbf{R}^{1/2} \in \mathbb C^{N \times N}$ denotes the correlation matrix for massive receiving antennas. $\widetilde{\mathbf{H}}_{k} \in \mathbb{C}^{N \times 1}$ is the fast fading coefficients with i.i.d. elements. Substituting Eq. \eqref{eqn:channel} into Eq. \eqref{eqn:rec_sig}, we have the following matrix form
\begin{equation}\label{eqn:rec_sig_mo}
\mathbf{y} = \sum\nolimits_{k=1}^{K} \sigma_k^{1/2} \mathbf{R}^{1/2} \widetilde{\mathbf{H}}_{k} s_k + \mathbf{n} = \mathbf{R}^{1/2} \mathbf{H} \mathbf{\Sigma}^{1/2} \mathbf{s} +\mathbf{n},
\end{equation}
where
\begin{equation}\label{eqn:para}
  \left\{
  \begin{aligned}
    \mathbf{H} &= [\widetilde{\mathbf{H}}_1 \ \widetilde{\mathbf{H}}_2 \ \ldots \ \widetilde{\mathbf{H}}_K],\\
    \mathbf{\Sigma}^{1/2} &= \textrm{diag} [\sigma_1^{1/2} \ \sigma_2^{1/2} \ \ldots \ \sigma_K^{1/2}],\\
    \mathbf{s} &= [s_1 \ s_2 \ \ldots \ s_K]^T.
  \end{aligned}
  \right.
\end{equation}

Without loss of generality, we assume the classical MMSE equalizer is applied for the uplink M-MIMO system, and the estimated symbol at the receiver side is thus given by
\begin{equation}\label{eqn:est}
\resizebox{.82\hsize}{!}{$
\begin{aligned}
\hat{\mathbf{s}} =& \left(\underbrace{\left(\mathbf{R}^{1/2} \mathbf{H} \mathbf{\Sigma}^{1/2} \right)^H \mathbf{R}^{1/2} \mathbf{H} \mathbf{\Sigma}^{1/2} + \eta^2 \mathbf{I}_K}_{\mathbf{W}} \right)^{-1}\\
&\cdot\left(\mathbf{R}^{1/2} \mathbf{H} \mathbf{\Sigma}^{1/2} \right)^H \mathbf{y}.
\end{aligned}
$}
\end{equation}

The most challenging task in MMSE is to compute the inversion operation of the filtering matrix $\mathbf{W}$. In the traditional MIMO systems, since the number of received antennas $N$ is usually small, e.g. $N \leq 8$, the matrix inversion operation can be efficiently computed through the standard exact decomposition methods, such as QR-Gram Schmidt method \cite{singh2007vlsi}, QR-Givens Rotation method \cite{kim2008practical}, and Gauss-Jordan method \cite{arias2011suitable}. If we note that $\mathbf{W}$ is Hermitian, we can further apply Cholesky decomposition method to reduce the implementation complexity as shown in \cite{studer2011asic}. However, when the numbers of received antennas $N$ and of UTs $K$, become significant (e.g. $N \geq 64$ and $K \geq 30$), the brute-force implementation may not be able to suffer due to the high power consumption and the massive storage requirement\footnote{In general, the conventional exact inversion methods (including Cholesky decomposition) require complicated data-flow mechanisms with sequential computation order, and involve hardware operations such as square-roots and divisions. As a result, those methods consume expensive implementation costs in terms of area, power, and latency, especially when the active UTs are large and varying.}.

\subsection{Neumann Series Approximation}
As an alternative approach to compute the large-scale matrix inversion, the Neumann series expansion (NSE) \cite{gw1998matrix} takes the advantage of iterative structure to gradually improve the accuracy, which avoids the prohibitive power and storage cost for general exact decomposition methods \cite{feng:15}. For illustration purpose, we summarize the theoretical background in Lemma~\ref{lem:NS} and describe the benchmark implementation in what follows.
\begin{Lem}[Neumann series]\label{lem:NS} For a non-singular $M \times M$ matrix $\mathbf{P}$ with $\lim_{i \rightarrow \infty} \mathbf{P}^{i} = \mathbf{0}_{M}$, we have $(\mathbf{I}_M - \mathbf{P})$ is also non-singular and its inverse is given by
\begin{equation}
(\mathbf{I}_M - \mathbf{P})^{-1} = \sum\nolimits_{i=0}^{\infty} \mathbf{P}^{i}.
\end{equation}
\end{Lem}

Applying Lemma \ref{lem:NS} and with some mathematical manipulations, we can rewrite the inverse of $\mathbf{W}$ using NSE and obtain
$\mathbf{W}^{-1} = \sum_{i=0}^{\infty} \big(\mathbf{I}_{K} - \widehat{\mathbf{W}}^{-1} \mathbf{W}\big)^{i} \widehat{\mathbf{W}}^{-1} $, where $\widehat{\mathbf{W}}$ is an arbitrary matrix satisfying
\begin{equation}\label{equ:convergen}
\lim_{i \rightarrow \infty} \big(\mathbf{I}_{K} - \widehat{\mathbf{W}}^{-1} \mathbf{W}\big)^{i} = \mathbf{0}_{K}.
\end{equation}

To write them into the iterative form, define $\mathbf{W}^{-1}(k) = \sum_{i=0}^{k} \big(\mathbf{I}_{K} - \widehat{\mathbf{W}}^{-1} \mathbf{W}\big)^{i} \widehat{\mathbf{W}}^{-1} $. We have
\begin{equation}\label{equ:ite}
\left\{
\begin{aligned}
&\mathbf{W}^{-1}(1) =  \widehat{\mathbf{W}}^{-1},\\
&\mathbf{W}^{-1}(L+1) = \mathbf{\Theta} \mathbf{W}^{-1}(L) + \widehat{\mathbf{W}}^{-1},~L=1,2,\ldots
\end{aligned}
\right.
\end{equation}
where $\mathbf{\Theta} = \mathbf{I}_{K} - \widehat{\mathbf{W}}^{-1} \mathbf{W}$ is the multiplication coefficient.


The following assumptions are adopted through the rest of the paper. First, we generate the receiving correlation matrix $\mathbf{R}$ as $r_{lm}=(\zeta e^{j\omega_{lm}})^{|l-m|}~(l\leq m)$ and $r_{ml}=r_{lm}^*$, according to \cite{godana:2013}. Here $\zeta~(0\leq \zeta\leq 1)$ denotes the correlation coefficient between consecutive antennas and $\omega_{lm}$ denotes a given phase. Second, perfect channel-state information is assumed to be available at the receiver side. Third, we assume non-correlation among different UTs in the uplink.

\section{Low-Complexity Tridiagonal Matrix Approximation Algorithm}
In the conventional approach to deal with M-MIMO systems \cite{Marzetta:12}, we usually rely on the fact that $\lim_{N \rightarrow \infty}\frac{1}{N} (\mathbf{H}^{H} \mathbf{H}) = \mathbf{I}_K$ to simplify the filtering matrix $\mathbf{W}$ under the uncorrelated channel, e.g., $\mathbf{W} = N \mathbf{\Sigma}$ when $\mathbf{R} = \mathbf{I}_N$. The remaining channel inversion operation is rather straightforward. Even for general large value of $N$, the off-diagonal elements are relatively small and we can rely on $\widehat{\mathbf{W}} =  \mathbf{W}_{\textrm{dia}}=\mathrm{diag}_0{(\mathbf{W})}$ to efficiently compute $\mathbf{W}^{-1}$ using NSE \cite{Matthaiou:10,Rusek:13}. However, for correlated M-MIMO environment, there is still lack of systematical ways for efficient implementation. In this section, we first analyze the characteristics of correlated M-MIMO channels and then propose a tridiagonal approximation algorithm to improve the accuracy of NSE.

\subsection{Characteristics of Correlated M-MIMO Channels}
With the receiving antenna correlation, one of the most important issues is to obtain the statistical information of $\mathbf{W}$, from which we can analyze and possibly simplify the corresponding receiving structure using NSE. In what follows, we derive the statistical information of $\mathbf{W}$ with general receiving correlation $\mathbf{R}$ and summarize the main results in Theorem \ref{thm:cor}.

\begin{Thm} In the correlated MIMO environment with the receiving antenna correlation matrix $\mathbf{R}$, the filtering matrix, $\mathbf{W}$, has the following statistical property, where its mean and variance matrices are given by Eq.s (\ref{eqn:mean}) and (\ref{eqn:var}), respectively.
\begin{equation}\label{eqn:mean}
\mathbf{\Sigma}_{\mathbf{W}}  =  \mathbb{E} \{ \mathbf{W} \} = N \mathbf{\Sigma}.
\end{equation}

\begin{equation}\label{eqn:var}
\begin{aligned}
\mathbf{\Omega}_{\mathbf{W}}  &= \mathbb{E} \{ \left(\mathbf{W - \mathbf{\Sigma}_{\mathbf{W}}} \right) \left(\mathbf{W - \mathbf{\Sigma}_{\mathbf{W}}}\right)^{H}\}\\
 &= \|\mathbf{R}(\zeta)\|^2_{F}\mathbf{\Sigma}\mathbf{1}_{K}\mathbf{\Sigma}.
\end{aligned}
\end{equation}
\label{thm:cor}
\end{Thm}

\begin{Rem}
Please refer to Appendix \ref{pf:thm:cor} for the proof.
\end{Rem}

\begin{figure}[htbp]
\centering
\includegraphics[width=0.4\textwidth]{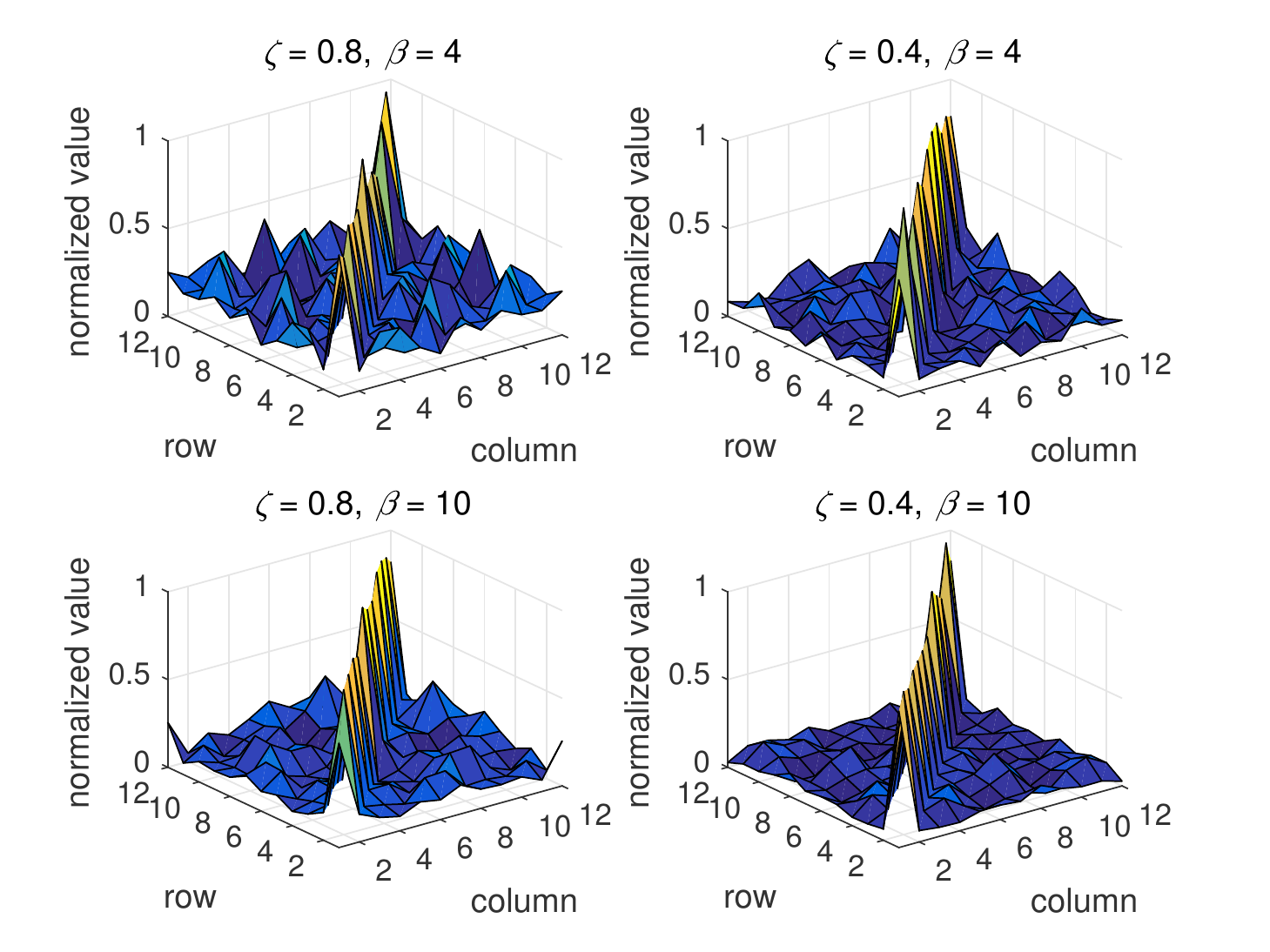}
\caption{The normalized distribution of $|w_{ij}|/\mathrm{max}|w_{ij}|$ with a fixed number of UTs, $K=12$. $\mathbf{W}$ tends to be strongly diagonally dominant in a favorable condition with large $\beta$ or small $\zeta$.}
\label{fig:mag_matrix}
\end{figure}

From Theorem \ref{thm:cor}, the channel variance matrix $\mathbf{\Omega}_{\mathbf{W}}$ has been greatly affected by the receiving correlation $\mathbf{R}$, whereas the statistical mean $\mathbf{\Sigma}_{\mathbf{W}} $ does not change if compared with uncorrelated case. In this case, the magnitude of off-diagonal elements in the filtering matrix $\mathbf{W}$ actually increases with the Frobenius norm \cite{golub:1996} of the receiving correlation matrix $\|\mathbf{R}(\zeta)\|_{F}$, and decreases with the number of antenna elements $N$, as shown in Fig. \ref{fig:mag_matrix}. The recent work has proven that a larger $\beta$ ($=N/K$) would be favorable propagation, so we next mainly analyze the effect of correlation on the property of $\mathbf{W}$ in this paper.

\begin{figure}[htbp]
\centering
\includegraphics[width=0.4\textwidth]{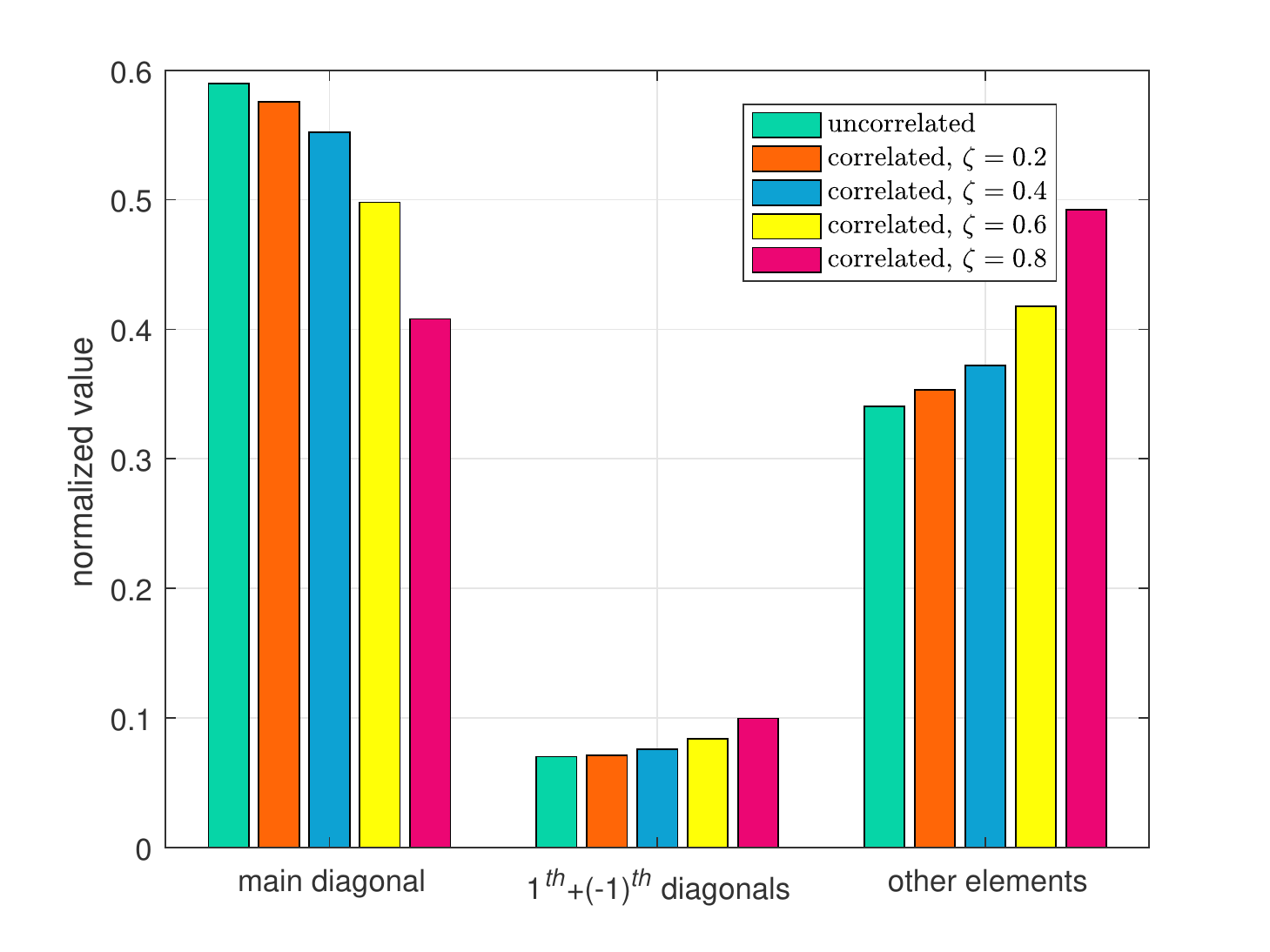}
\caption{Comparison of the normalized $l^1$-norm between main diagonal and other $k^{\text{th}}$ diagonals with a fixed ratio, $K=12$, $\beta=16$, for different propagation channels.}
\label{fig:dia_off}
\end{figure}

If we further plot the $l^1$-norm \cite{golub:1996} of diagonal elements versus off diagonal elements as shown in Fig. \ref{fig:dia_off}, we can observe that the $l^1$-norm of the main ($0^{\text{th}}$) diagonal decreases with the correlation coefficient $\zeta$, whereas $l^1$-norms of other $k^{\text{th}}$ $(1 \leq |k|\leq K-1)$ diagonals increase. This is partially because of the ever increasing ratio $\|\mathbf{\Omega}_{\mathbf{W}}\|_F / \|\mathbf{\Sigma}_{\mathbf{W}}\|_F$. The conventional NSE way to compute $\mathbf{W}^{-1}$ relying on $\widehat{\mathbf{W}} =  \mathbf{W}_{\textrm{dia}}$ may result in inaccuracy and slow convergence.

\begin{figure}[htbp]
\centering
\includegraphics[width=0.4\textwidth]{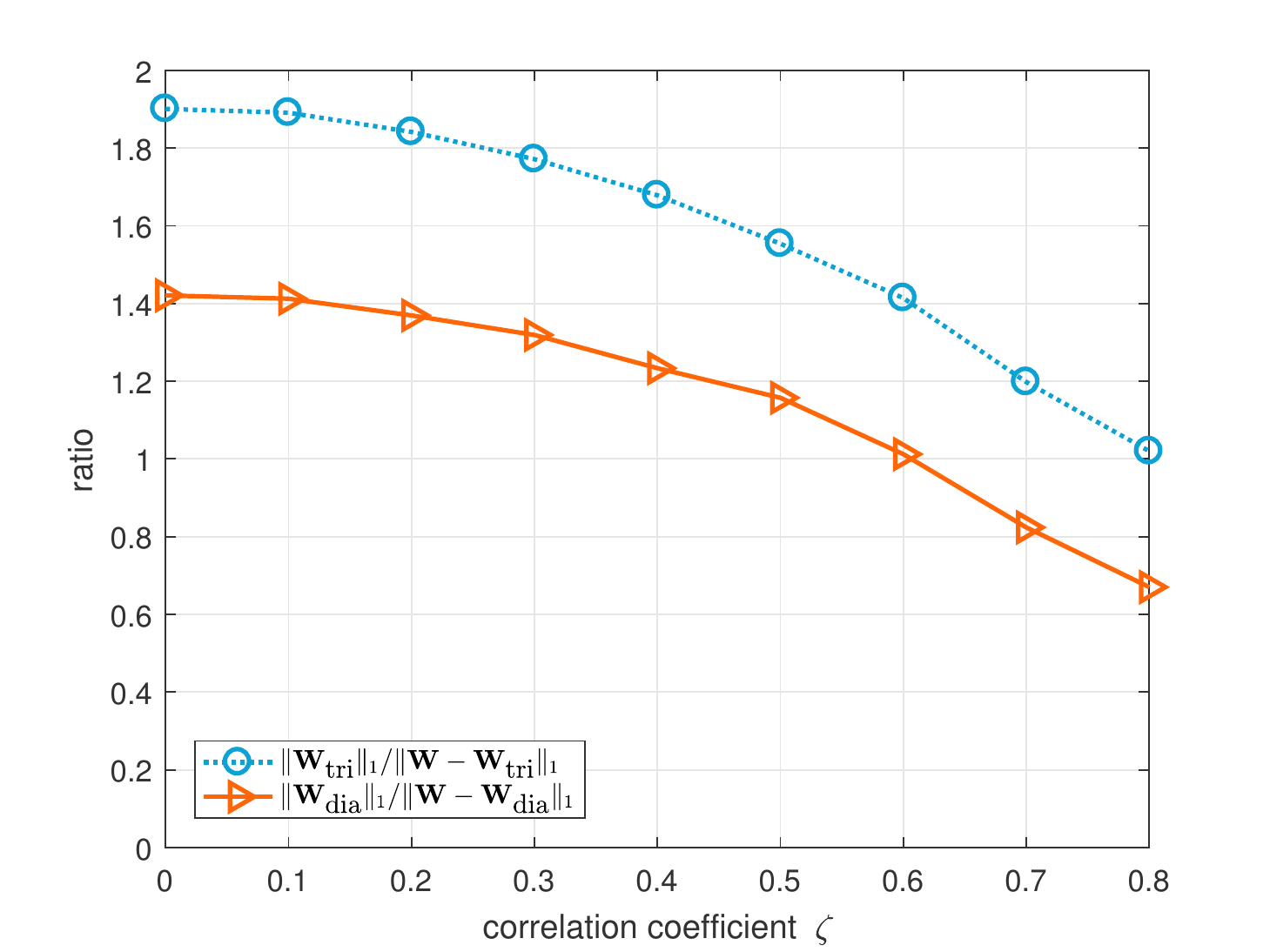}
\caption{\textcolor{black}{Comparison of  $\|\mathbf{W}_{\textrm{tri}}\|_{1}/\|\mathbf{W}-\mathbf{W}_{\textrm{tri}}\|_{1}$ and $\|\mathbf{W}_{\textrm{dia}}\|_{1}/\|\mathbf{W}-\mathbf{W}_{\textrm{dia}}\|_{1}$, with $K=12$, $\beta=16$.}}
\label{fig:tri-dig}
\end{figure}

As shown in Fig. \ref{fig:tri-dig}, the ratio of $\|\mathbf{W}_{\textrm{tri}}\|_{1}/\|\mathbf{W}-\mathbf{W}_{\textrm{tri}}\|_{1}$, where $\mathbf{W}_{\textrm{tri}}=\sum_{k=-1}^{1}{\textrm{diag}_{k}(\mathbf{W})}$, is at least $30\%$ larger than the ratio $\|\mathbf{W}_{\textrm{dia}}\|_{1}/\|\mathbf{W}-\mathbf{W}_{\textrm{dia}}\|_{1}$ in various correlation condition. So NSE relying on $\widehat{\mathbf{W}} =  \mathbf{W}_{\textrm{tri}}$ would help to improve the convergence stability and speed according to \cite{gw1998matrix}.

\subsection{Error Analysis}\label{sec:err}
In practical M-MIMO systems, the number of NSE terms $L$ cannot be very large because of the considerations on latency and complexity. In this case, we next analyze the error resulted from the truncated polynomial, $\mathbf{W}^{-1}(L)=\sum_{n=0}^{L-1}{\mathbf{\Theta}^{n}\widehat{\mathbf{W}}^{-1}}$ , for MMSE estimation in the massive systems. We define the approximation error matrix as $\mathbf{\Delta}_{\mathbf{W}|L}=\mathbf{W}^{-1}-\mathbf{W}^{-1}(L)$, and compute the estimated symbol at the receiver side with the $L$-term approximation $\mathbf{W}^{-1}(L)$, i.e.,
\begin{equation}
\begin{aligned}
\mathbf{s}_{\mathbf{W}|L}=&\mathbf{W}^{-1}(L)(\mathbf{R}^{1/2}\mathbf{H}\mathbf{\Sigma}^{1/2})^{H}\mathbf{y}\\
=&\hat{\mathbf{s}}-\mathbf{\Delta}_{\mathbf{W}|L}(\mathbf{R}^{1/2}\mathbf{H}\mathbf{\Sigma}^{1/2})^{H} \mathbf{y},
\end{aligned}
\end{equation}
where $\mathbf{\Delta}_{\mathbf{W}|L}(\mathbf{R}^{1/2}\mathbf{H}\mathbf{\Sigma}^{1/2})^{H} \mathbf{y}$ is the \textit{residual estimation error}. We denote the square of $l^2$-norm of \textit{residual estimation error} by $\Phi=\|\mathbf{\Delta}_{\mathbf{W}|L}(\mathbf{R}^{1/2}\mathbf{H}\mathbf{\Sigma}^{1/2})^{H} \mathbf{y}\|_{2}^{2}$ and bound $\Phi$ by:
\begin{equation}
\Phi=\left\|\mathbf{\Theta}^{L}\hat{\mathbf{s}}\right\|^{2}_{2}\leq
\|\mathbf{\Theta}^{L}\|_{F}^{2}\|\hat{\mathbf{s}}\|_{2}.
\label{equ:phi}
\end{equation}

In M-MIMO systems, let the number of receive antennas $N$ grow large while keeping the number of transmit antennas $K$ constant, we further assume the uncorrelated noise is averaged out, which is corresponding to \cite{Rusek:13}. In this case, the expect of $\Phi$ can be derived as:
\begin{equation}
\mathbb{E}\{\Phi\}=\mathbb{E}\{\left\|\mathbf{\Theta}^{L}\right\|^{2}_{F}\}.
\label{equ:ephi}
\end{equation}
\begin{Rem}
Please refer to Appendix \ref{pf:thm:phi} for the proof.
\end{Rem}

From Eq.s (\ref{equ:phi}) and (\ref{equ:ephi}), the size of $\Phi$ is determined by $\|\mathbf{\Theta}^{L}\|_{F}$. \textcolor{black}{Considering} the convergence condition in NSE, i.e., $\|\mathbf{\Theta}\|_{F}< 1$, one intuitive method to reduce $\Phi$ is to realize a large $L$-term approximation. As a limit case, the $\Phi$ caused by the approximation error $\mathbf{\Delta}_{\mathbf{W}|L}$ is zero when $L\rightarrow\infty$.

\begin{figure}[ht]
\centering
\includegraphics[width=0.42\textwidth]{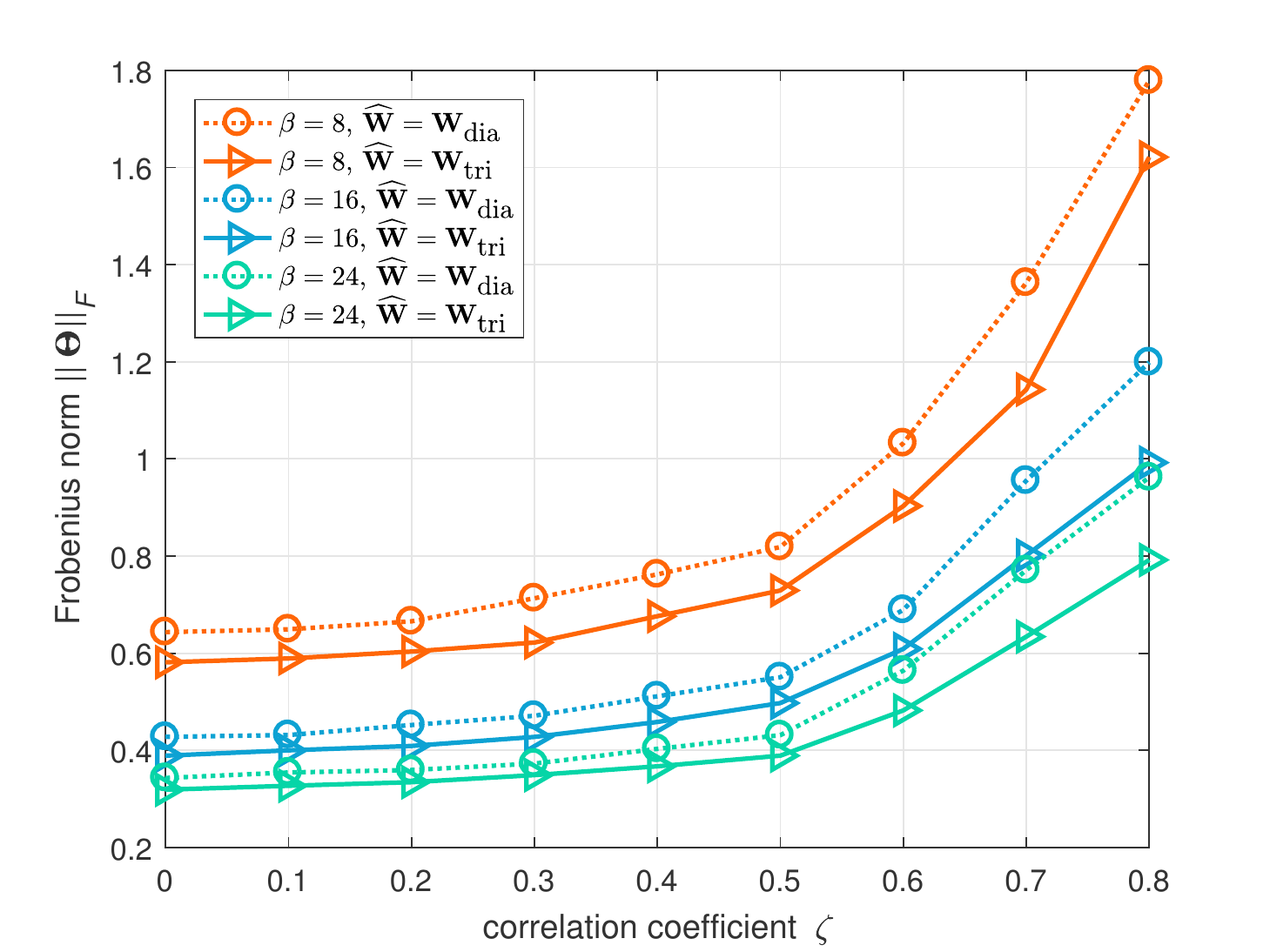}
\caption{\textcolor{black}{$\|\Theta\|_{F}$ when $\widehat{\mathbf{W}}=\mathbf{W}_{\textrm{tri}}$ and $\mathbf{W}_{\textrm{dia}}$, $K=12$.}}
\label{fig:tri-dig-2norm}
\end{figure}

Another method to reduce $\Phi$ is to decrease the consecutive norm of the inner matrix, $\mathbf{\Theta}$. Concretely, in view of $\mathbf{\Theta}=(\mathbf{I}_{K}-\widehat{\mathbf{W}}^{-1}\mathbf{W})$, $\|\mathbf{\Theta}\|_{F}$ can reduce faster if the initial matrix $\widehat{\mathbf{W}}$ is more close to the filtering matrix $\mathbf{W}$. As shown in Fig. \ref{fig:tri-dig-2norm}, the probability that the inner matrix $\mathbf{\Theta}$ has a smaller Frobenius norm when using $\widehat{\mathbf{W}}=\mathbf{W}_{\textrm{tri}}$ is always higher than the method relying on $\widehat{\mathbf{W}}=\mathbf{W}_{\textrm{dia}}$, which means an improved convergence speed and stability. Moreover, the scenarios where the correlation is high, i.e., $\zeta\geq 0.6$, with a small $\beta$ may not satisfy the convergence condition of NSE, which means a much unfavorable propagation environment. Thus, in the remainder of this paper, we mainly concern the propagation environment where $\zeta \leq 0.6$.

\subsection{Tridiagonal Approximation Algorithm}
To further improve the stability and convergence for the NSE method accommodating less favorable propagation environment, a valid method is to acquire more off-diagonal elements, i.e., $ \widehat{\mathbf{W}}_{l} =\textrm{diag}_{0}(\mathbf{W})+\sum_{|a|=1}^{l}{\textrm{diag}_{a}(\mathbf{W})} $. Though setting up $\widehat{\mathbf{W}}_{l}$ requires more computations as compared to choosing $\widehat{\mathbf{W}} =  \mathbf{W}_{\textrm{dia}}$, the overall complexity can be expected to reduce because of the improved convergence rate, which in-turn would require smaller iteration number $L$ in Eq. (\ref{equ:ite}). Therefore, one would enjoy a better performance/complexity tradeoff. Shown in Fig. \ref{fig:tri-dig}, it is obvious that $\|\mathbf{W}_{\textrm{tri}}\|_{1}$ occupies a dominated proportion in magnitude, whereas other $\|\textrm{diag}_{a}(\mathbf{W})\|_{1}$s account for relatively small proportion. Then we initialize the matrix as $\widehat{\mathbf{W}}=\mathbf{W}_{\textrm{tri}}$. For lower complexity and better convergence for NSE, a modified method to compute the approximate inversion of $\mathbf{W}_{\textrm{tri}}$ is developed. It should be noted that a tri-band matrix inversion for pre-coding was proposed in \cite{lund2014iscas}.

Inspired by existing literature \cite{Moawwad:11}, the straightforward inversion of the tridiagonal matrix $\mathbf{W}_{\textrm{tri}}$ could be carried out as follows. For better explanation, we denote $\mathbf{W}_{\textrm{tri}}$ by
\begin{equation}\label{eqn:tri}
\resizebox{.88\hsize}{!}{$
\mathbf{W}_{\textrm{tri}}=\left(
    \begin{array}{ccccccc}
      b_1 & c_1 &  &  &  &  & \\
      a_2 & b_2 & c_2 &  &  &  &\\
       & . & . & . &  &  &\\
       &  & . & . & . &  &\\
       & &  &  & a_{K-1} & b_{K-1} & c_{K-1} \\
       & &  &  &  & a_K & b_K \\
    \end{array}
  \right).
  $}
\end{equation}

The inverse matrix $\mathbf{W}_{\textrm{tri}}^{-1}=\{\phi_{ij}\}$ can be given by
\begin{equation}\label{eqn:diag}
\phi_{ii}=\frac{1}{b_i-a_ic_{i-1}\frac{z_{i-2}}{z_{i-1}}-a_{i+1}c_i\frac{y_{i+2}}{y_{i+1}}},
\end{equation}
where $i=1,2,\ldots,K$, $(a_1,c_K)=(0,0)$. Also
\begin{equation}\label{eqn:offdiag}
\phi_{ij}=
\left\{
\begin{aligned}
&-c_i\frac{z_{i-1}}{z_i}\phi_{(i+1)j},~ {i < j}, \\
&-a_i\frac{y_{i+1}}{y_i}\phi_{(i-1)j},~ {i > j}.
\end{aligned}
\right.
\end{equation}

Other parameters are given by the second-order linear recurrences as follows:
\begin{equation}\label{eqn:iterme}
\resizebox{.88\hsize}{!}{$
  \left\{
  \begin{aligned}
    y_j&=b_jy_{j+1}-a_{j+1}c_jy_{j+2},~j=K-1,K-2,\ldots,1,\\
    z_i&=b_iz_{i-1}-a_ic_{i-1}z_{i-2},~i=2,3,\ldots,K,\\
  \end{aligned}
  \right.
  $}
\end{equation}
where the initialization conditions are $(z_0,z_1)=(1,b_1)$ and $(y_{K+1},y_K)=(1,b_K)$.

One critical issue of this straightforward inversion operation is the prohibitive complexity involved, which is $\mathcal{O}(K^2)$. Note that the complexity here is estimated by counting the number of complex multiplications. For M-MIMO systems with considerable large $K$, the resulting high complexity may hinder the application of such inversion method in real systems.
We again note that the filtering matrix $\mathbf{W}$ is Hermitian. Then vector $\mathbf{a}=[a_2 \ a_3 \ \ldots \ a_n]$ is the conjugate transpose of $\mathbf{c}=[c_1 \ c_2 \ \ldots \ c_{n-1}]$. Since the inverse of a Hermitian matrix is still Hermitian, only half of the required entries have to be calculated with the sophisticated equations of Eq.s (\ref{eqn:diag}) and (\ref{eqn:offdiag}). The other half could be obtained conveniently by simple conjugate operation. Admittedly, the Hermitian property does help to reduce about half complexity, the complexity stays in the same order and fails to meet our initial expectation for introducing NSE. For a better balance of performance and complexity, especially in correlated massive MIMO systems, further simplifications are expected.

To this end, in what follows, we first decompose the straightforward tridiagonal approach into two steps: 1) \textbf{Step 1}, which includes Eq.s (\ref{eqn:diag}) and (\ref{eqn:offdiag}), mainly deals with the entry calculation; 2) \textbf{Step 2}, which includes Eq. (\ref{eqn:iterme}), mainly deals with the parameter calculation. Then, stepwise optimizations are carried out by looking into details of each step. Statistical property of the filtering matrix $\mathbf{W}$ has been made full use to achieve a better performance/complexity tradeoff.

\subsubsection{Simplification for Step 1}\label{sec:step1}
Since the inverse of a tridiagonal matrix is a full matrix, the complexity of Step 1 is  $\mathcal{O}(K^2)$. However, in correlated M-MIMO channels, $l^1$-norm of the main diagonal for $\mathbf{W}_{\textrm{tri}}$ is higher than that of the secondary diagonals, which implies further simplification.

Let us only consider the lower triangular part of $\mathbf{W}_{\textrm{tri}}^{-1}$, i.e., $i < j$.  According to the formula $\phi_{ij}=-c_i\frac{z_{i-1}}{z_i}\phi_{(i+1)j}$, the calculation starts with the main diagonal, then recursively computes the off-diagonal entries in a column-wise manner. The entries of vector $\mathbf{p}=\{-a_i\frac{z_{i-1}}{z_i}\}$ will be multiplied recursively. If we understand how the magnitude of entries of $\mathbf{p}$ change with channels, we will be able to depict the off-diagonal entries' attenuation accordingly. We evaluate the overall magnitude of $\mathbf{p}$ by its mean $p=\overline{\mathbf{p}}$. The tendency of $p$ with correlation coefficient $\zeta$ changing is given in Table \ref{tab:itww}.
\begin{table}[ht]
    \tabcolsep 1mm
    \renewcommand{\arraystretch}{1.2}
    \footnotesize
    \caption{The tendency of \textcolor{black}{$p$} with correlation coefficient $\zeta$ changing}
    \begin{center}
    \begin{tabular}{ c || c | c | c | c | c | c }
    \Xhline{1.0pt}
     $\zeta$ & $0$ & $0.2$ & $0.4$ & $0.6$ & $0.8$& $1$  \\
    \hline
     $p$  & $0.054$ & $0.058$ & $0.071$ & $0.101$ & $0.194$& $0.506$  \\
     \Xhline{1.0pt}
    \end{tabular}\label{tab:itww}
    \end{center}
    \end{table}

It is observed that when $\zeta < 0.6$, $p$ stays less than $0.1$, which indicates that for channels with $\zeta < 0.6$, the magnitudes of outer diagonals shrink at the scale of $10\%$ layer by layer. In other words, the dominated portion of $\mathbf{W}_{\textrm{tri}}^{-1}$ fall into the region between two secondary diagonals. The rest part of $\mathbf{W}_{\textrm{tri}}^{-1}$, whose magnitude is less than $1\%$ of the main diagonal, can be omitted as a result. Thus, the complexity will be effectively reduced to $\mathcal{O}(K)$.

\subsubsection{Simplification for Step 2}\label{sec:step2}
We recall that in correlated M-MIMO channels, $l^1$-norm of the main diagonal for $\mathbf{W}_{\textrm{tri}}$ is higher than that of the secondary diagonals. Therefore, the second term at the right-hand-side of the following equation can be reasonably removed:
\begin{equation*}
y_j=b_jy_{j+1}-a_{j+1}c_jy_{j+2},~j=K-1, K-2, \ldots, 1.
\end{equation*}
Then, we have the following approximation:
\begin{equation}\label{eqn:app}
y_j=b_jy_{j+1},~j=K-1,K-2,\ldots,1.
\end{equation}

With this approximation, the formula for calculating the main diagonal of $\mathbf{W}_{\textrm{tri}}^{-1}$ can be simplified as:
\begin{equation}
\tilde{\phi}_{ii}=\frac{1}{b_i-|a_i|^2\frac{z_{i-2}}{z_{i-1}}-\frac{|a_{i+1}|^2}{b_{i+1}}}.
\end{equation}

For previous calculation in Eq. (\ref{eqn:iterme}), two sequences $\{y_j\}$ and $\{z_i\}$ are obtained in ascending and descending orders of $i$ or $j$, respectively. Thanks to the approximation, there is no more need to calculate $\{y_j\}$ and the processing latency has been successfully halved.

To evaluate the performance degradation, numerical results are given in Table \ref{tab:ee}, by denoting the error percentage as $\frac{|\|\tilde{\phi}_{ii}\|_{1}-\|\phi_{ii}\|_{1}|}{\|\phi_{ii}\|_{1}}$. Since observation shows that It is observed that when $\zeta < 0.6$ the error percentage stays below $1\%$, it is safe to introduce this approximation within this region. For clearer denotation, we use $\phi_{ii}$ instead of $\tilde{\phi}_{ii}$ in following.
\begin{table}[ht]
    \tabcolsep 1mm
    \renewcommand{\arraystretch}{1.2}
    \footnotesize
    \caption{Error percentage with proposed approximation when $\zeta$ changes}
    \begin{center}
    \begin{tabular}{c || c | c | c | c | c | c }
    \Xhline{1.0pt}
     $\zeta$ & $0$ & $0.2$ & $0.4$ & $0.6$ & $0.8$& $1$  \\
    \hline
     Error ($\%$)  & $0.256$ & $0.252$ & $0.408$ & $0.931$ & $17.557$& $212.54$  \\
     \Xhline{1.0pt}
    \end{tabular}\label{tab:ee}
    \end{center}
    \end{table}

With the aforementioned simplification steps, now the new algorithm to calculate $\mathbf{W}_{\textrm{tri}}^{-1}$ is given by Alg. \ref{alg:triapp} as follows:
\begin{algorithm}
\caption{Proposed Tridiagonal Matrix Inversion}\label{alg:triapp}
\begin{algorithmic}[1]
\REQUIRE {$\mathbf{W}=\{w_{ij}\}$, $1 \leq i,j\leq K$, $0\leq |j-i|\leq 1$}\\
\STATE \textbf{Initialize:}  $\phi_{11}=(w_{11}-\frac{w_{21} \cdot w_{12}}{w_{22}})^{-1}$\\
\FOR{$i=2:K$}
\STATE $ratio=\frac{w_{i(i-1)}}{w_{(i-1)(i-1)}}$\label{1:first:1}
\STATE $\phi_{i(i-1)}=-ratio$
\STATE $w_{ii}=w_{ii}-ratio \times w_{(i-1)i}$\label{1:first:2}
\STATE $w_{i(i-1)}=w_{ii}-\frac{w_{(i+1)i} \times w_{i(i+1)}}{w_{(i+1)(i+1)}}$\label{1:second:1}
\ENDFOR
\FOR{$i=2:K$}
\STATE $\phi_{ii}=\frac{1}{w_{i(i-1)}}$\label{1:second:2}
\STATE $\phi_{i(i-1)}=\frac{\phi_{i(i-1)}}{w_{i(i-1)}}$
\STATE $\phi_{(i-1)i}=\phi^{*}_{i(i-1)}$
\ENDFOR
\ENSURE {$\mathbf{W}_{\textrm{tri}}^{-1}=\{\phi_{ij}\}$, $1 \leq i\leq K$, $0\leq |j-i|\leq 1$}
\end{algorithmic}
\end{algorithm}
\begin{Rem}
\textcolor{black}{Please refer to Appendix \ref{pf:thm:all} for the derivation.}
\end{Rem}

The statistical property of the filtering matrix $\mathbf{W}$ inspires us to reduce the computational complexity to calculate $\mathbf{W}_{\textrm{tri}}^{-1}$. Though $\mathbf{W}_{\textrm{tri}}^{-1}$ is a full matrix, the entries outside the three central diagonals are notably small in magnitude. Therefore, we mainly focus on the three dominating diagonals, i.e., $\mathbf{W}_{\textrm{tri}}^{-1}\approx\textrm{diag}_{0}(\mathbf{W}_{\textrm{tri}}^{-1})+\textrm{diag}_{1}(\mathbf{W}_{\textrm{tri}}^{-1})+\textrm{diag}_{-1}(\mathbf{W}_{\textrm{tri}}^{-1})$. For the purpose of parallel computation, some units are modified accordingly. The resulting hardware overhead is usually small compared to the matrix-matrix multiplications. Furthermore, we only directly calculate the lower-triangular part of the Hermitian matrix.

We denote this tridiagonal matrix inversion approximation by TMA. Simulation results are in Section \ref{sec:mat}. The detailed hardware implementation is shown in Section \ref{sec:fpga} as follows.

\section{Hardware-Friendly Implementation}\label{sec:fpga}
\subsection{Efficient Pipelining Structure}
The TMA formula can be written as follows,
\begin{eqnarray}
\resizebox{.92\hsize}{!}{$
\begin{cases}
\mathbf{W}^{-1}(1) =  \mathbf{W}^{-1}_{\textrm{dia/tri}},\\
\mathbf{W}^{-1}(L+1) = \mathbf{\Theta} \mathbf{W}^{-1}(L) + \mathbf{W}^{-1}_{\textrm{dia/tri}} , \ L=1,2,\ldots
\end{cases},
\label{equ:iter}
$}
\end{eqnarray}
where $\mathbf{\Theta} = \big(\mathbf{I}_{K} - \mathbf{W}^{-1}_{\textrm{dia/tri}} \mathbf{W}\big)$ is the multiplication coefficient. According to the linear detection formula Eq. (\ref{eqn:est}), general framework of hardware can be designed as in Fig. \ref{fig:ARC}. The framework mainly consists of the following parts:

\subsubsection{Orange Block}
It is the pre-processing module (PM), where $\mathbf{\Gamma}=\mathbf{R}^{1/2} \mathbf{H} \mathbf{\Sigma}^{1/2}$, and $\eta^2$ is the normalized variance of white Gaussian noise (AWGN) with zero mean. The output of PM is $\mathbf{W}$, which can be split as $\mathbf{W}=\mathbf{X}+\mathbf{E}$. For the diagonal Neumann series (DNS) method, matrix $\mathbf{X}=\mathbf{W}_{\mathrm{dia}}$ includes all the diagonal elements of $\mathbf{W}$; whereas for the TMA, matrix $\mathbf{X}=\mathbf{W}_{\textrm{tri}}$ only includes the tridiagonal elements in $\mathbf{W}$.

\subsubsection{Blue Block}
It is the main-computing module (MM), which uses TMA to calculate $\mathbf{W}^{-1}(L)$ with $L$ iterations. The main structure of this part is IIR-like, whose iterations can be controlled to realize different accuracy requirements.

\subsubsection{Green Block}
It is the estimation module (EM), which computes $\hat{\mathbf{s}}=\mathbf{W}^{-1}(L)\cdot\hat{\mathbf{y}}$, where $\hat{\mathbf{y}}=\mathbf{\Gamma}^{H}\cdot\mathbf{y}$.

\begin{figure}[ht]
\centering
\includegraphics[width=0.42
\textwidth]{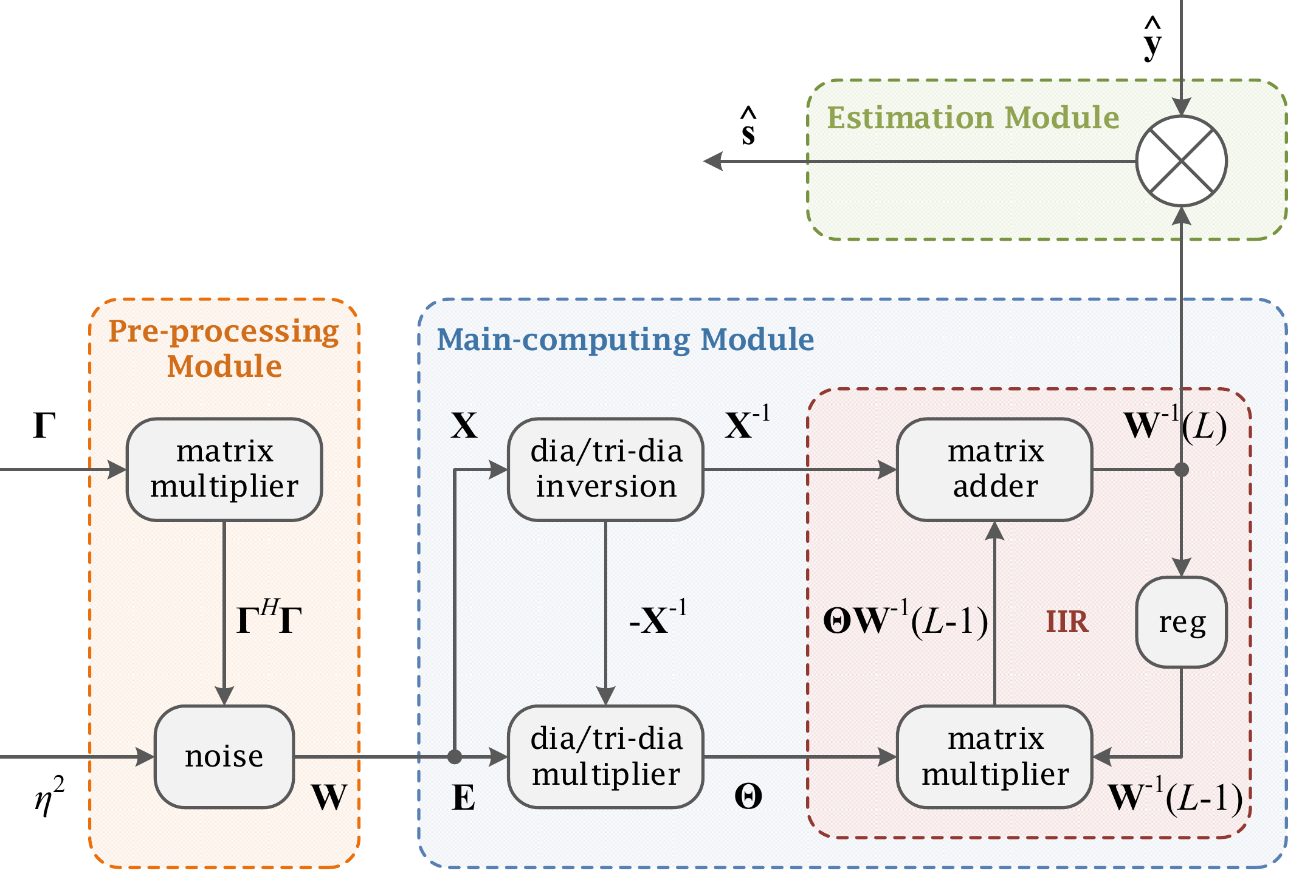}
\caption{Proposed linear detection architecture with TMA.}
\label{fig:ARC}
\end{figure}

\subsection{Diagonal/Tri-Diagonal Matrix Inversion}
This module calculates the inversion matrix of $\mathbf{X}$, which is a diagonal matrix in the DNS scheme. So the inversion matrix of $\mathbf{W}_{\mathrm{dia}}$ is used to compute the reciprocals of diagonal elements. In order to compute $\mathbf{W}_{\mathrm{dia}}^{-1}$, a reciprocal unit based on look-up table (LUT), which is particularly suitable for FPGA implementation, is employed.

The tridiagonal matrix inversion has been simplified in Alg. \ref{alg:triapp}, and the correspondent hardware design can be seen in Fig. \ref{fig:pl1}. This architecture is realized in a pipelining way, in which each clock $i$ increases from $1$ to $K$ and $w_{ii}$ and $w_{(i-1)i}$ change accordingly. A switch is employed for scheduling: 1) when $i=1$, the switch is open and the output is $\phi_{11}$; 2) when $2\leq i\leq K$, the switch is closed and outputs $\phi_{ii}$ and $\phi_{i(i-1)}$. Different blocks have different functions.

\subsubsection{Red Block}
It operates iteratively to realize States \ref{1:first:1} and \ref{1:first:2} in Alg. \ref{alg:triapp}.

\subsubsection{Blue Block}
It completes States \ref{1:second:1} and \ref{1:second:2} to compute the main diagonal elements of $\mathbf{W}_{\textrm{tri}}^{-1}$.

\subsubsection{Green Block}
It computes the secondary diagonal elements of lower triangle. For $\mathbf{W}_{\textrm{tri}}$ is Hermitian, $\mathbf{W}_{\textrm{tri}}^{-1}$ is Hermitian as well. With conjugate operation we can get the entire matrix $\mathbf{W}_{\textrm{tri}}^{-1}$.

The inversion needs $2$ real adders, $5$ complex multipliers, and $3$ LUTs. The latency of this module is $K$ clocks, where $K$ is the size of $\mathbf{W}$.
\begin{figure}[thb]
\centering
\includegraphics[width=0.42
\textwidth]{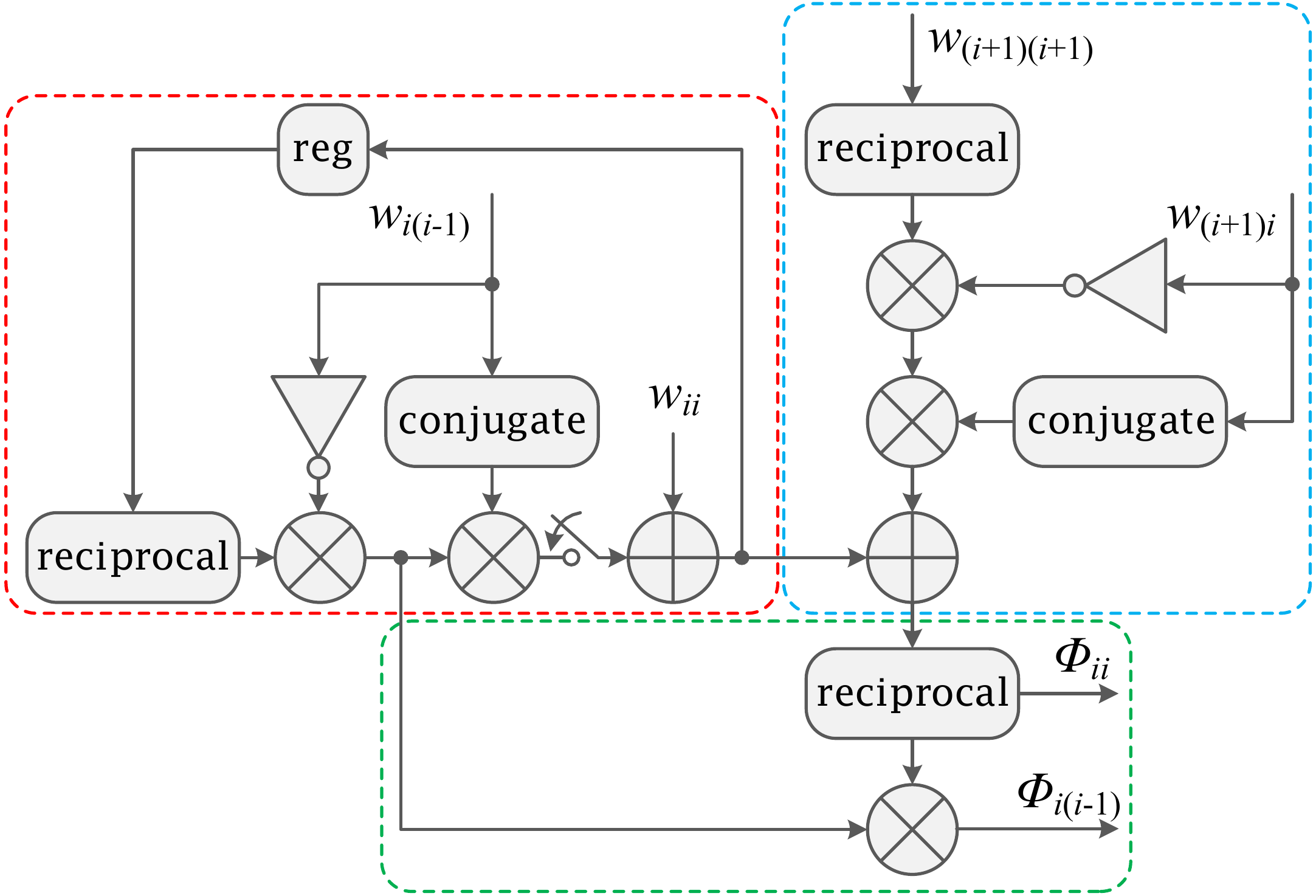}
\caption{Architecture of tridiagonal inversion based on Alg. \ref{alg:triapp}.}
\label{fig:pl1}
\end{figure}

\begin{figure}[thb]
\centering
\includegraphics[width=0.42\textwidth]{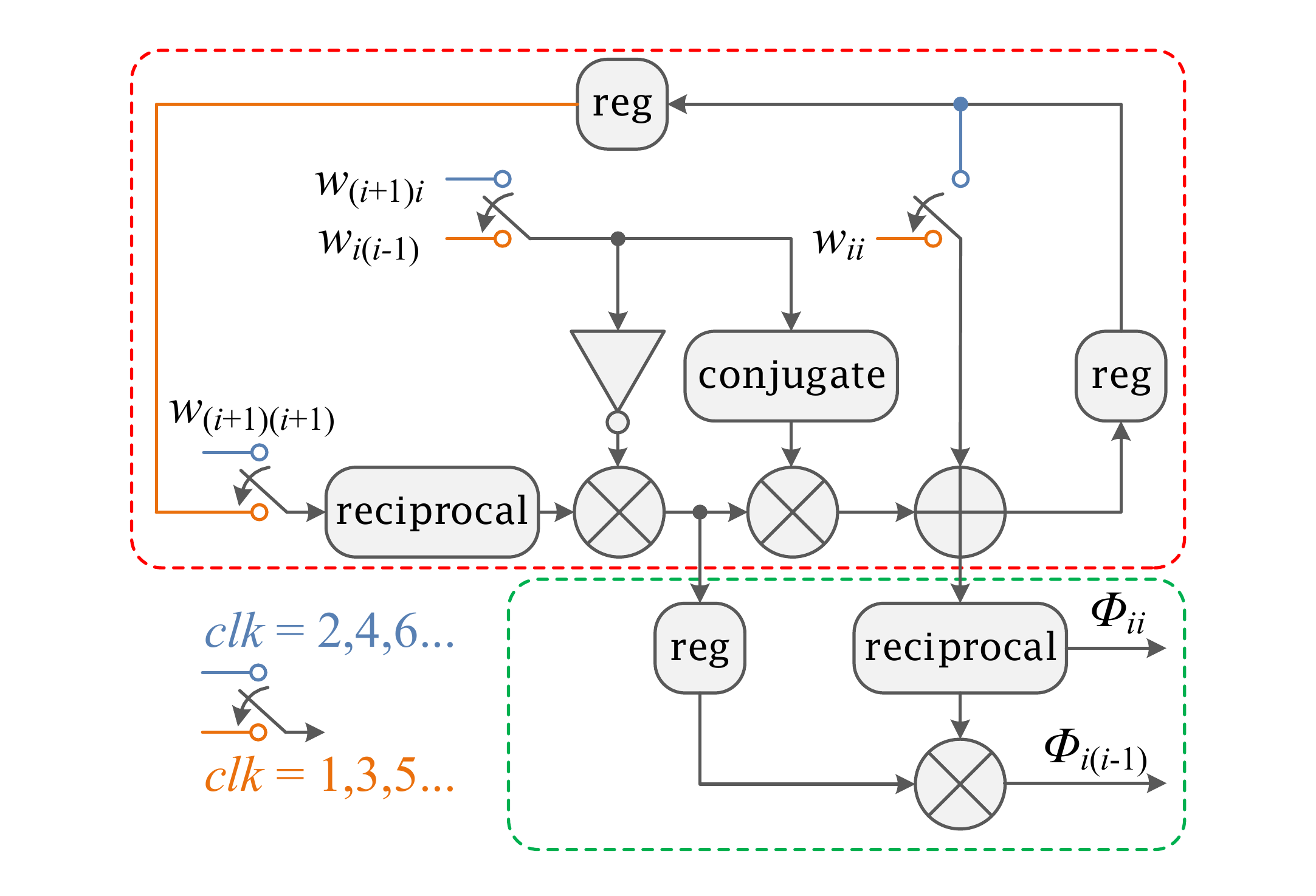}
\caption{Architecture of tridiagonal inversion based on Alg. \ref{alg:triapp2}.}
\label{fig:pl2}
\end{figure}

Alg. \ref{alg:triapp} can be converted to Alg. \ref{alg:triapp2} for lower implementation complexity. We combine the common hardware of two identical algorithm states, which could save half of the hardware cost. It is noted that States \ref{1:first:1} and \ref{1:first:2} in Alg. \ref{alg:triapp} share the same iterative operation as States \ref{1:second:1} and \ref{1:second:2}. Therefore, the common hardware can be reused. The resulting architecture also works in a pipelining way but $i$ increases one every two clocks. Now, the work done in one clock previously needs two clocks to complete, and latency doubles. The folded architecture is in Fig. \ref{fig:pl2}, which executes States \ref{first:1} and \ref{first:2} (or States \ref{second:1} and \ref{second:2}) in Alg. \ref{alg:triapp2} when $clk = 1,3,5...$ (or $clk = 2,4,6...$). It needs $1$ real adder, $3$ complex multipliers, $2$ LUTs, and $2K$ clocks.

\begin{algorithm}
\caption{Proposed Tridiagonal Matrix Approximation}
\begin{algorithmic}[1]
\REQUIRE {$\mathbf{W}=\{w_{ij}\}$, $1 \leq i,j\leq K$, $0\leq |j-i|\leq 1$}\\
\STATE \textbf{Initialize:}  $\phi_{11}=(w_{11}-\frac{w_{21} \cdot w_{12}}{w_{22}})^{-1}$\\
\FOR{$i=2:K$}
\STATE $ratio1=\frac{w_{i(i-1)}}{w_{(i-1)(i-1)}}$ \label{first:1}
\STATE $\phi_{i(i-1)}=-ratio1$ \label{second:1}
\STATE $w_{ii}=w_{ii}-ratio1\times w_{(i-1)i}$ \label{first:2}
\STATE $ratio2=\frac{w_{i(i+1)}}{w_{(i+1)(i+1)}}$ \label{second:2}
\STATE $w_{i(i-1)}=w_{ii}-ratio2\times w_{(i+1)i}$
\ENDFOR
\FOR{$i=2:K$}
\STATE $\phi_{ii}=\frac{1}{w_{i(i-1)}}$
\STATE $\phi_{i(i-1)}=\frac{\phi_{i(i-1)}}{w_{i(i-1)}}$
\STATE $\phi_{(i-1)i}=\phi^{*}_{i(i-1)}$
\ENDFOR
\ENSURE {$\mathbf{W}_{\textrm{tri}}^{-1}=\phi_{ij}$, $1 \leq i\leq K$, $0\leq |j-i|\leq 1$}
\end{algorithmic}
\label{alg:triapp2}
\end{algorithm}

\begin{Rem}\label{rem:4}
  Architectures in both Fig.s \ref{fig:pl1} and \ref{fig:pl2} can take care both diagonal and tri-diagonal matrix inversions.
\end{Rem}

\subsection{Diagonal/Tri-Diagonal Multiplier}
\begin{figure}[ht]
\centering
\includegraphics[width=0.42\textwidth]{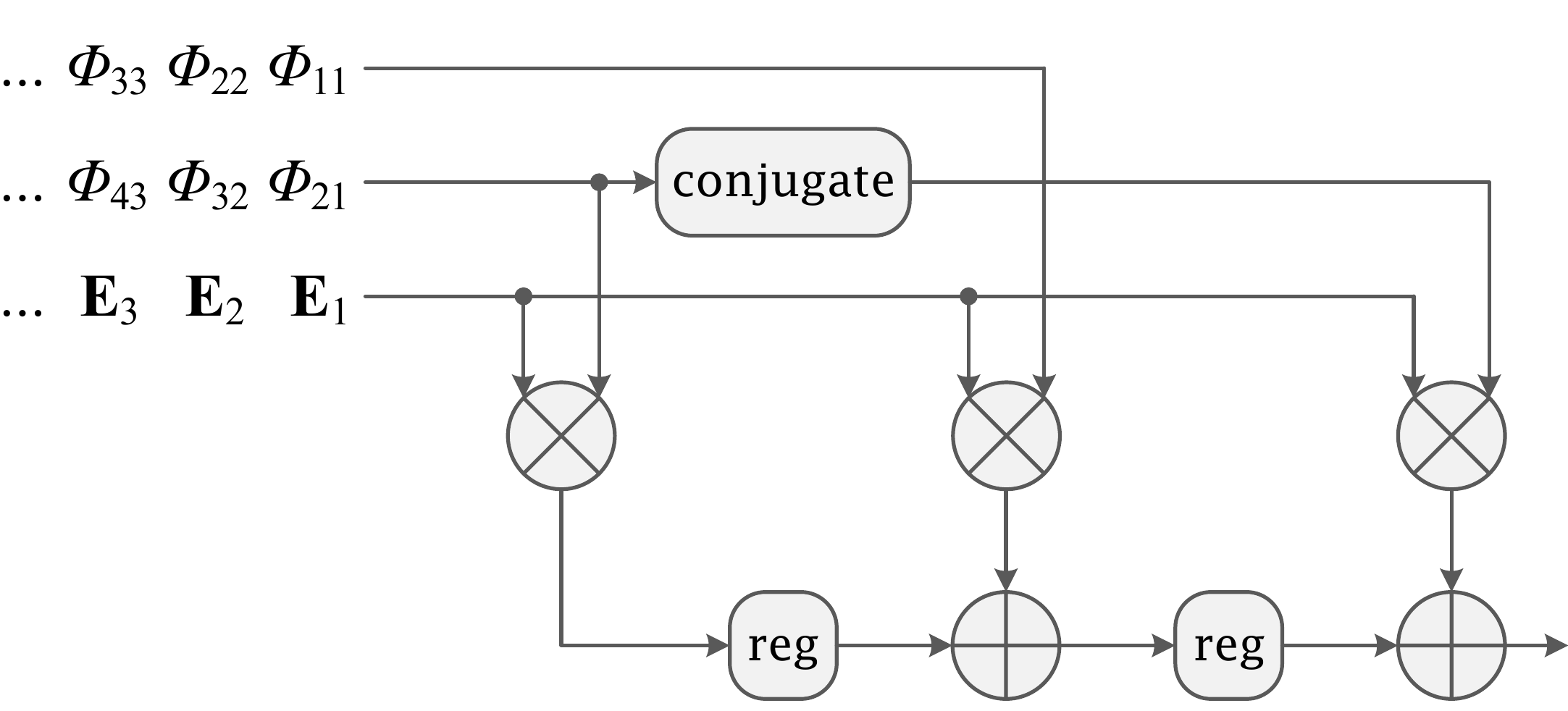}
\caption{Architecture of tridiagonal matrix multiplier, where $\phi_{ij}$ is corresponding element of the matrix and $\mathbf{E}_i$ denotes corresponding row in the matrix.}
\label{fig:tri}
\end{figure}

This module calculates matrix $\mathbf{\Theta}$, which is the product of matrices $\mathbf{X}^{-1}$ and $\mathbf{E}=\mathbf{W}-\mathbf{X}$. No matter $\mathbf{X}$ is a diagonal matrix $\mathbf{W}_{\mathrm{dia}}$ or tridiagonal matrix $\mathbf{W}_{\textrm{tri}}$, the channel hardening property makes the multiplication module simple. To calculate $\mathbf{W}_{\mathrm{dia}}^{-1}(\mathbf{W}-\mathbf{W}_{\mathrm{dia}})$, every diagonal element of $\mathbf{W}_{\mathrm{dia}}^{-1}$ is multiplied with the corresponding row of $(\mathbf{W}-\mathbf{W}_{\mathrm{dia}})$. Furthermore, when calculating $\mathbf{W}_{\textrm{tri}}^{-1}(\mathbf{W}-\mathbf{W}_{\textrm{tri}})$, three lines of diagonal elements needs to product corresponding row of the latter matrix, and using registers to add three sets of results in correct combination. The hardware design can be seen in Fig. \ref{fig:tri}. By using $3K$ complex multipliers, $2K$ registers and $2K$ complex adders, the multiplication between tridiagonal matrix and any matrix can be realized. Meanwhile, when we input the main diagonal elements into the first set of multiplications, the product between diagonal matrix and any matrix can be realized as well. Therefore, we have:

\begin{Rem}\label{rem:5}
  Architecture in Fig. \ref{fig:tri} can implement multiplication of diagonal or tri-diagonal matrix with any matrix.
\end{Rem}

\subsection{Neumann Series Iteration Module}
This module is IIR-like, including a \emph{matrix multiplier} and a \emph{matrix adder}, which implements the matrix approximation inverse. \textcolor{black}{With $\mathbf{X}^{-1}$ and $\mathbf{\Theta}$ inputted, it outputs $\mathbf{W}^{-1}(L) = \mathbf{\Theta} \mathbf{W}^{-1}(L-1) + \mathbf{X}^{-1}$ in the $L^{\text{th}}$ iteration.} By controlling the iteration number, different accuracies can be achieved.

\emph{Matrix multiplier} is implemented by a systolic array. For $\mathbf{W}^{-1}(L)$ is Hermitian, once we calculate the lower-triangular part, the upper-triangular part can be obtained by conjugating. The systolic array has $(1+K)K/2$ complex adders and $(1+K)K/2$ complex multipliers. For \textcolor{black}{$\mathbf{\Gamma}^{H}\mathbf{\Gamma}$} is also Hermitian, the \emph{matrix multipliers} in PM and MM can be multiplexed with one lower-triangular systolic array. \emph{Matrix adder} needs one real adder for diagonal matrix addition, and two complex adders and one real adder for tridiagonal addition.
\begin{Rem}
Please refer to Appendix \ref{pf:thm:her} for the proof that $\mathbf{W}^{-1}(L)$ is Hermitian.
\end{Rem}

\begin{Rem}
  Remarks \ref{rem:4} and \ref{rem:5} indicate that the design in Fig. \ref{fig:ARC} can implement detection based on either DNS or TMA. \
\end{Rem}

\subsection{Complexity Analysis}
In this section, complexity comparison is given in terms of numbers of real adders and real multipliers. One complex adder includes two real adders, and one complex multiplier includes four real multipliers and two real adders. Then, the complexity of the proposed detector is:

\begin{Lem}
For one $K$-user TMA detection, the numbers of real adders ($RA$) and real multipliers ($RM$) are,
\begin{equation}
\left\{
\begin{aligned}
    RA_{DNS}&=2K^2+8K+1,\\
    RA_{TMA}&=2K^2+16K+12,\\
    RM_{DNS}&=2K^2+10K,\\
    RM_{TMA}&=2K^2+18K+12.
\end{aligned}
    \right.
\end{equation}
\end{Lem}

\begin{IEEEproof}
PM module mainly involves a low-triangular systolic array, which has $K^2/2+K/2$ complex adders and $K^2/2+K/2$ complex multipliers. The total cost is $2K^2+2K$ real adders and $2K^2+2K$ real multipliers.

Sharing the low-triangular systolic array with PM, MM becomes the main matrix inversion part: 1) for DNS, it costs $1$ real adders and $K$ complex multipliers ($2K+1$ real adders and $4K$ real multipliers in total), 2) for TMA, it costs $2$ real adders, $2K+2$ complex adders, and $3K+3$ complex multipliers ($10K+12$ real adders and $12K+12$ real multipliers in total).

EM module only involves a vector multiplier, which costs $K$ complex adders and $K$ complex multipliers ($4K$ real adders and $4K$ real multipliers in total).
\end{IEEEproof}

The complexity comparison of Cholesky decomposition (CD) \cite{feng:15,xiao:15}, DNS, and TMA is in Table \ref{tab:comple}.
\begin{table}[h]
\tabcolsep 1mm
\renewcommand{\arraystretch}{1.3}
\footnotesize
\setlength{\tabcolsep}{1.1pt}
\caption{Complexity comparison on different detection methods}
\begin{center}
\resizebox{!}{0.9cm}
{    \begin{tabular}{ c || c | c | c }
    \Xhline{1.0pt}
    Method & Latency (clocks) & Adder \# & Multiplier \#  \\
    \hline
    \rowcolor{mygray}
     CD & $ N+4K+5$ & $2K^3/3+4K/3$ & $2K^3+K^2+K/3$  \\
     DNS & $L(K+1)+N+2K-4$ & $2K^2+8K+1$ & $2K^2+10K$  \\
    \rowcolor{mygray}
     TMA  & $L(K+1)+N+2K-2$ & $2K^2+16K+12$ & $2K^2+18K+12$  \\
    \Xhline{1.0pt}
\end{tabular}\label{tab:comple}}
\end{center}
\end{table}

\subsection{Timing Analysis}\label{sec:time}
When $\mathbf{H} \in \mathbb{C}^{N \times K} $, the timing analyses is as follows:

\begin{Lem}
For one $K$ users, $N$ antennas MIMO system, the latency for TMA detection is,
\begin{equation}
L_{TMA}=L(K+1)+N+2K-2.
\end{equation}
\end{Lem}

\begin{IEEEproof}
Using the lower-triangular systolic array, PM generates its output from Clock $N$ to $(N+2K-2)$. In systolic array, when $w_{ij}$ is calculated in Clock $d$, $w_{(i+1)j}$ and $w_{i(j+1)}$ are calculated in Clock $d+1$. For $w_{11}$ is obtained in Clock $N$, $w_{KK}$ is obtained in Clock $(N+2K-2)$.

To calculate $\mathbf{W_{\textrm{tri}}^{-1}}$, each $\phi_{ii}$ needs $2$ clock cycles by Alg. \ref{alg:triapp2}, which fits the output frequency of $w_{ii}$. Thus, $\phi_{11}$ is obtained in Clock $(N+2)$, and $\phi_{KK}$ in Clock $(N+2K)$.

According to the rule of systolic array, $\mathbf{W}^{-1}(L)$ ($L>1$) is calculated from Clock $L(K+1)+N-1$ to $L(K+1)+N+2K-3$. After pipelining, the final estimation vector $\mathbf{\hat{s}}$ completes in Clock $L(K+1)+N+2K-2$.
\end{IEEEproof}

\begin{Lem}
For one $K$ users $N$ antennas MIMO system, the latency for DNS detection is,
\begin{equation}
L_{DNS}=L(K+1)+N+2K-4.
\end{equation}
\end{Lem}

\begin{IEEEproof}
Compared to TMA, \emph{dia inversion} requires one less clock for each $\phi_{ii}$. \emph{Dia multiplier} needs one less clock as well. Thus, the latency reduces by $2$.
\end{IEEEproof}

Since the latency is mainly spent for the iterative multiplication, as the iteration number $L$ grows, the latency becomes longer and throughput drops. Based on numerical results in Section \ref{sec:mat}, different $L$'s are set for different $\zeta$'s and $\beta$'s.

\section{Architecture Analysis}\label{sec:sce5}
\subsection{Numerical Results}\label{sec:mat}
We propose the Neumann series approximation to reduce the complexity of MIMO linear detection method. However, different iterations lead to different latencies, if $L$ becomes too large, the latency will be too high and result in low throughput. However, when $L$ is not big enough, the approximate result will not reach the required accuracy. Here, the accuracy metric is defined as: the performance degradation is less than $0.3$ dB compared with the Cholesky decomposition based exact MMSE detection when $\text{SNR}=10^{-3}$. Different sets of $\beta=N/K$ and correlated coefficients $\zeta$ result in quite different iteration counts.
\textcolor{black}{We compare on the iteration counts for diagonal scheme (DNS), tridiagonal scheme (TNS), and simplified tridiagonal scheme (TMA).}
\begin{table}[hb]
    \tabcolsep 1mm
    \renewcommand{\arraystretch}{1.2}
    \footnotesize
    \caption{Comparison on iteration counts for different $\zeta$'s and $\beta$'s}
    \begin{center}
    \begin{tabular}{ c || c | c | c | c | c | c }

    \Xhline{1.0pt}
    \multirow{2}{*}{Method} &\multicolumn{3}{c |}{$\beta=8$} &\multicolumn{3}{c}{$\beta=16$}\\
    \cline{2-7}& $\zeta=0$ & $\zeta=0.3$ & $\zeta=0.5$ & $\zeta=0$ &$\zeta=0.3$ & $\zeta=0.5$\\
    \hline
     DNS & $7$ & $13$ & $28$ & $4$ & $5$& $8$  \\
    \hline
     TMA  & $6$ & $11$ & $14$ & $3$ & $4$& $5$  \\
    \hline
    \Xhline{1.0pt}
    \end{tabular}\label{tab:it}
    \end{center}
    \end{table}
\begin{figure}[ht]
\centering
\includegraphics[width=0.42
\textwidth]{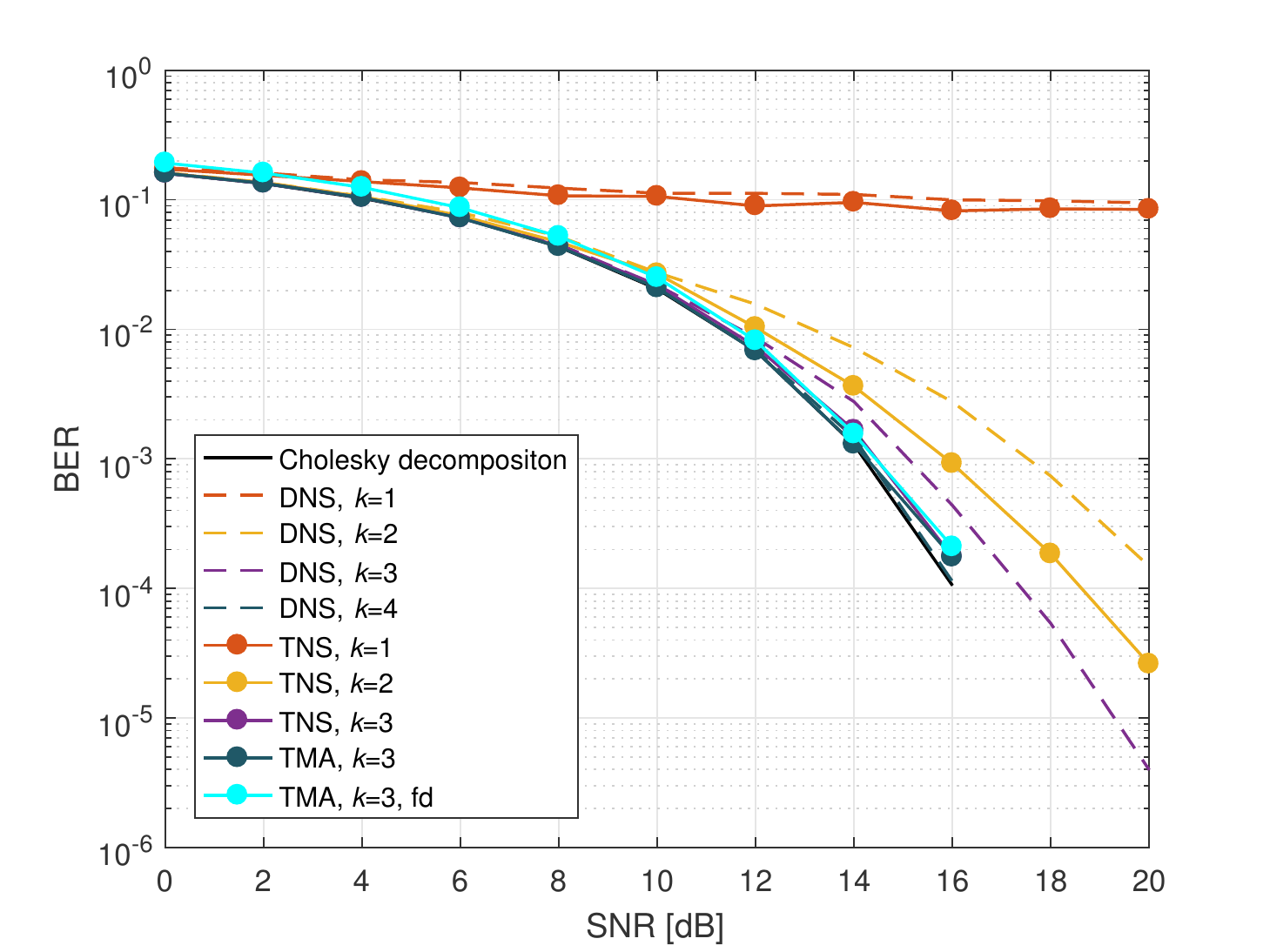}
\caption{\textcolor{black}{BER comparison among exact MMSE method, DNS, TNS, and TMA. Set $\beta=N/K=128/8=16$ in $\zeta=0$ channel by $64$-QAM.}}
\label{fig:r0}
\end{figure}
\begin{figure}[ht]
\centering
\includegraphics[width=0.42
\textwidth]{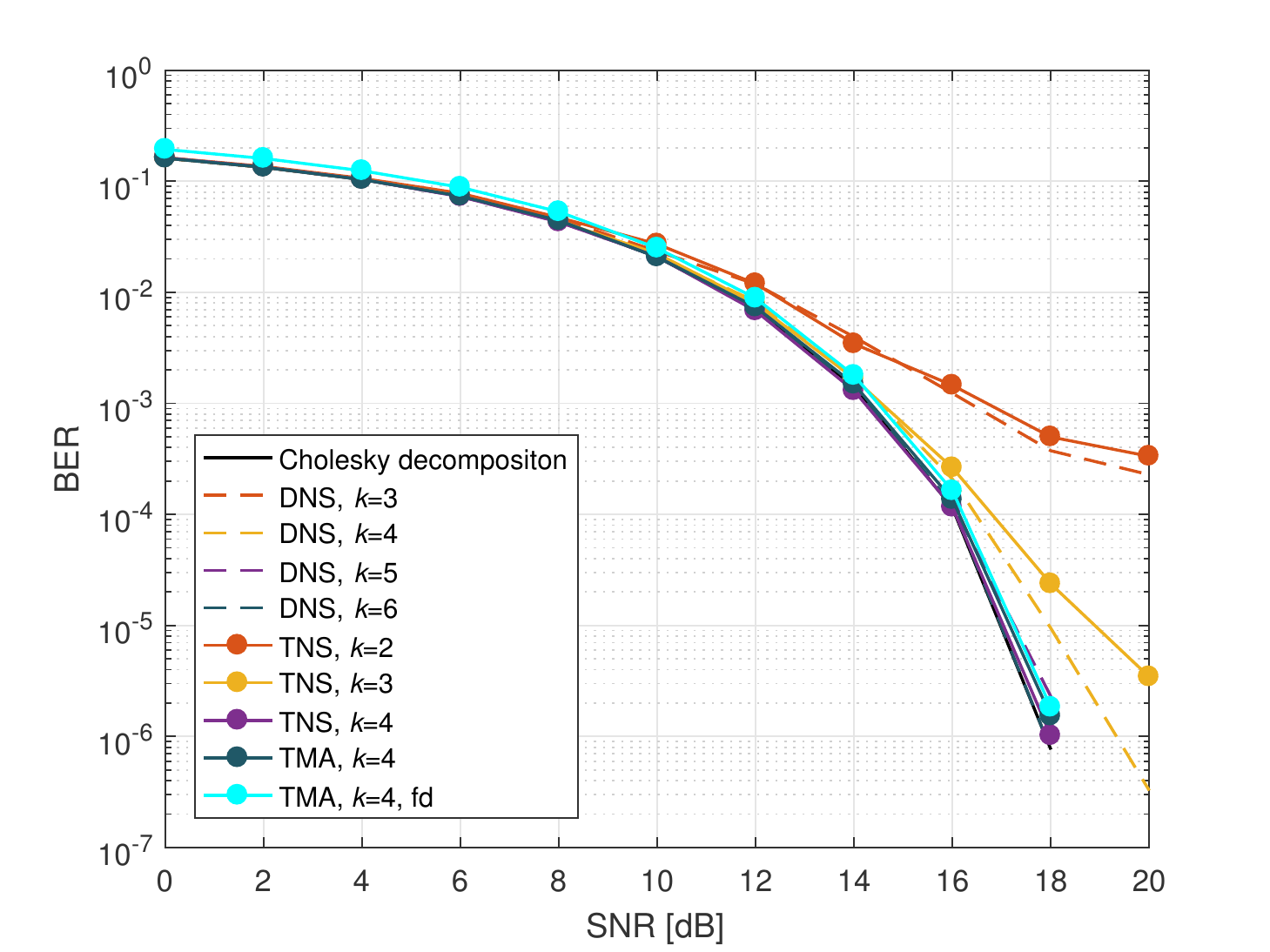}
\caption{\textcolor{black}{BER comparison among exact MMSE method, DNS, TNS, and TMA. Set $\beta=N/K=128/8=16$ in $\zeta=0.3$ channel by $64$-QAM.}}
\label{fig:r4}
\end{figure}
\begin{figure}[ht]
\centering
\includegraphics[width=0.42
\textwidth]{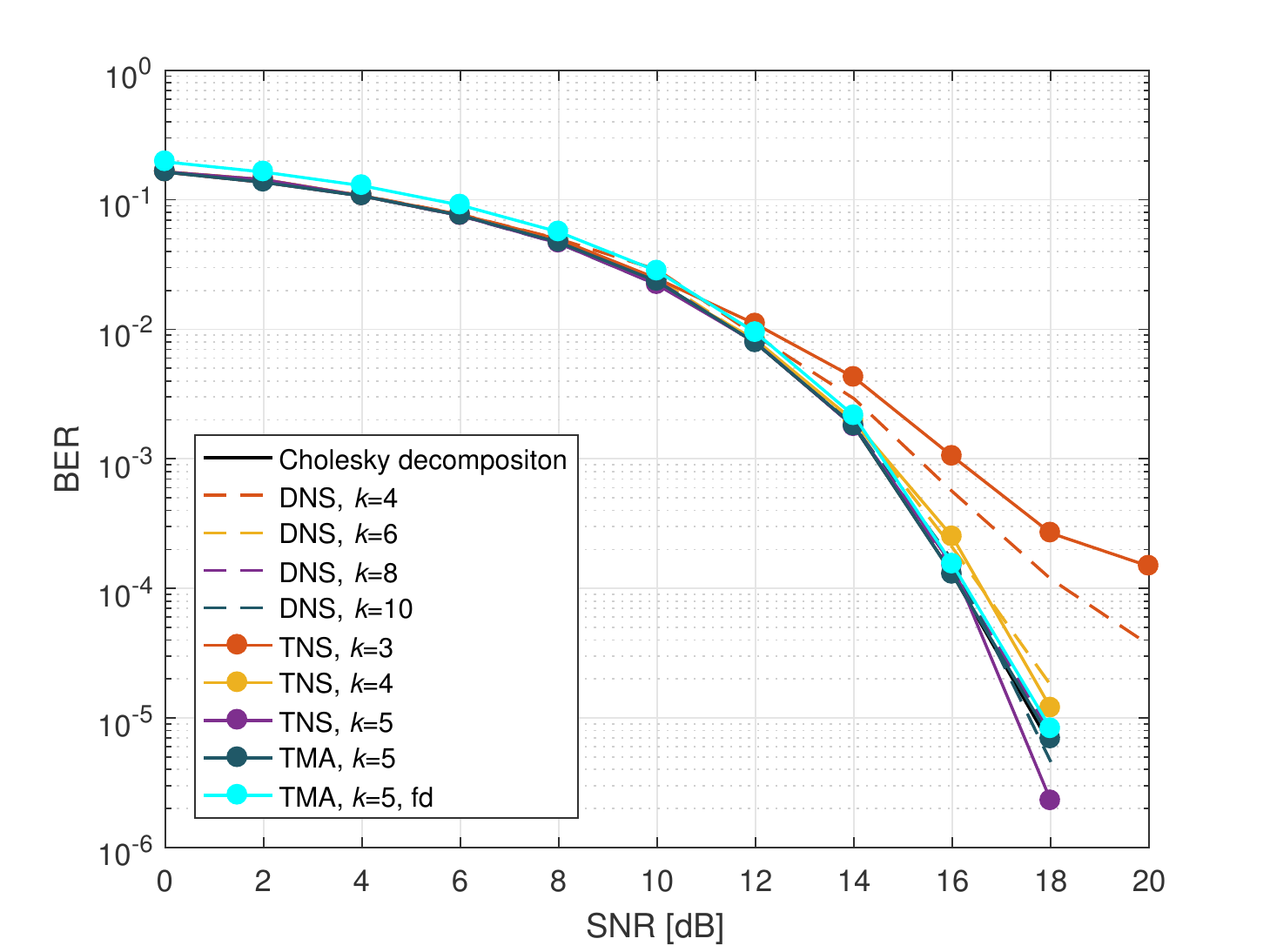}
\caption{\textcolor{black}{BER comparison among exact MMSE method, DNS, TNS, and TMA. Set $\beta=N/K=128/8=16$ in $\zeta=0.5$ channel by $64$-QAM.}}
\label{fig:r8}
\end{figure}
\begin{figure}[ht]
\centering
\includegraphics[width=0.42
\textwidth]{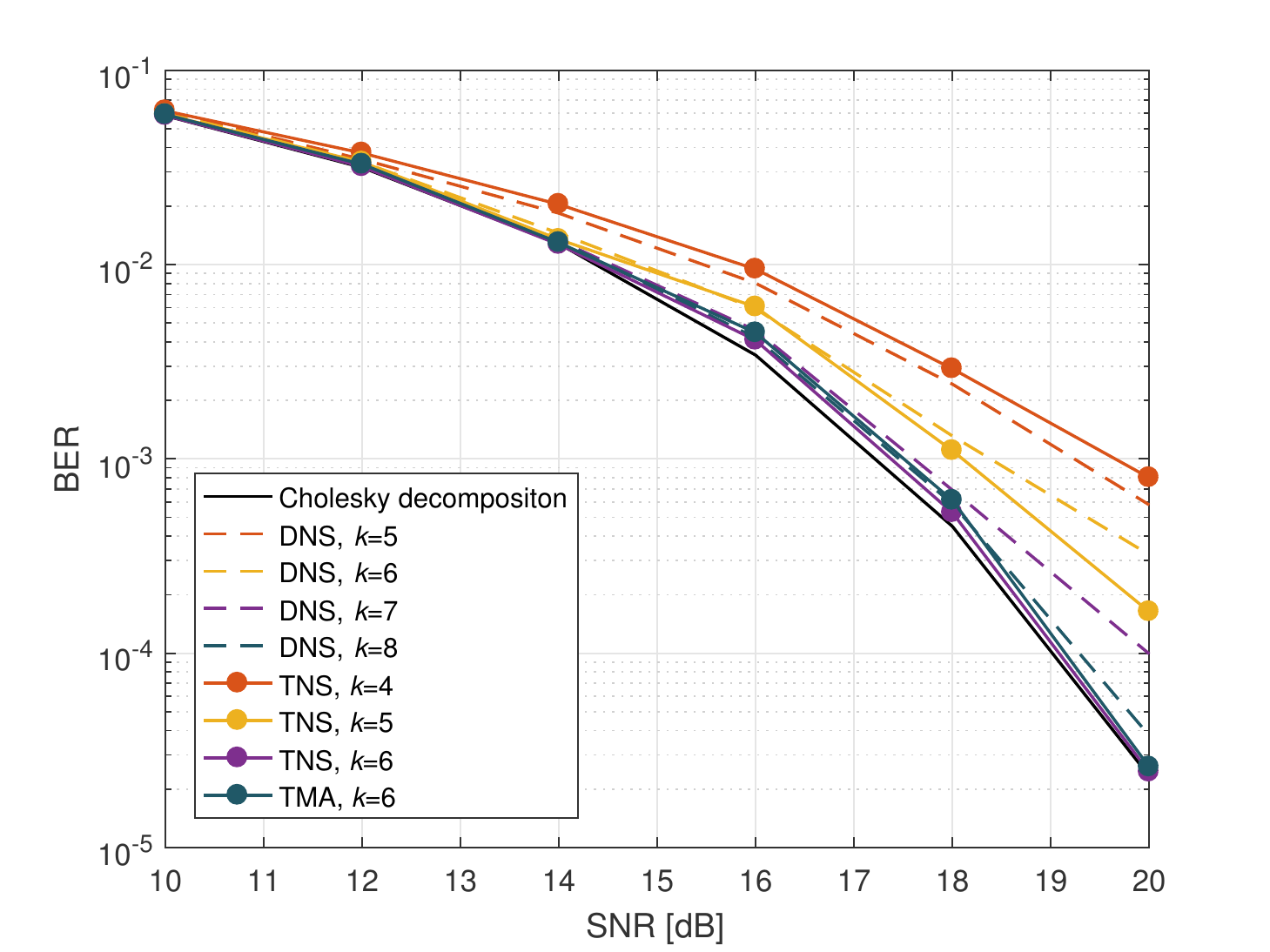}
\caption{\textcolor{black}{BER comparison among exact MMSE method, DNS, TNS, and TMA. Set $\beta=N/K=128/16=8$ in $\zeta=0$ channel by $64$-QAM.}}
\label{fig:r_0}
\end{figure}
\begin{figure}[ht]
\centering
\includegraphics[width=0.42
\textwidth]{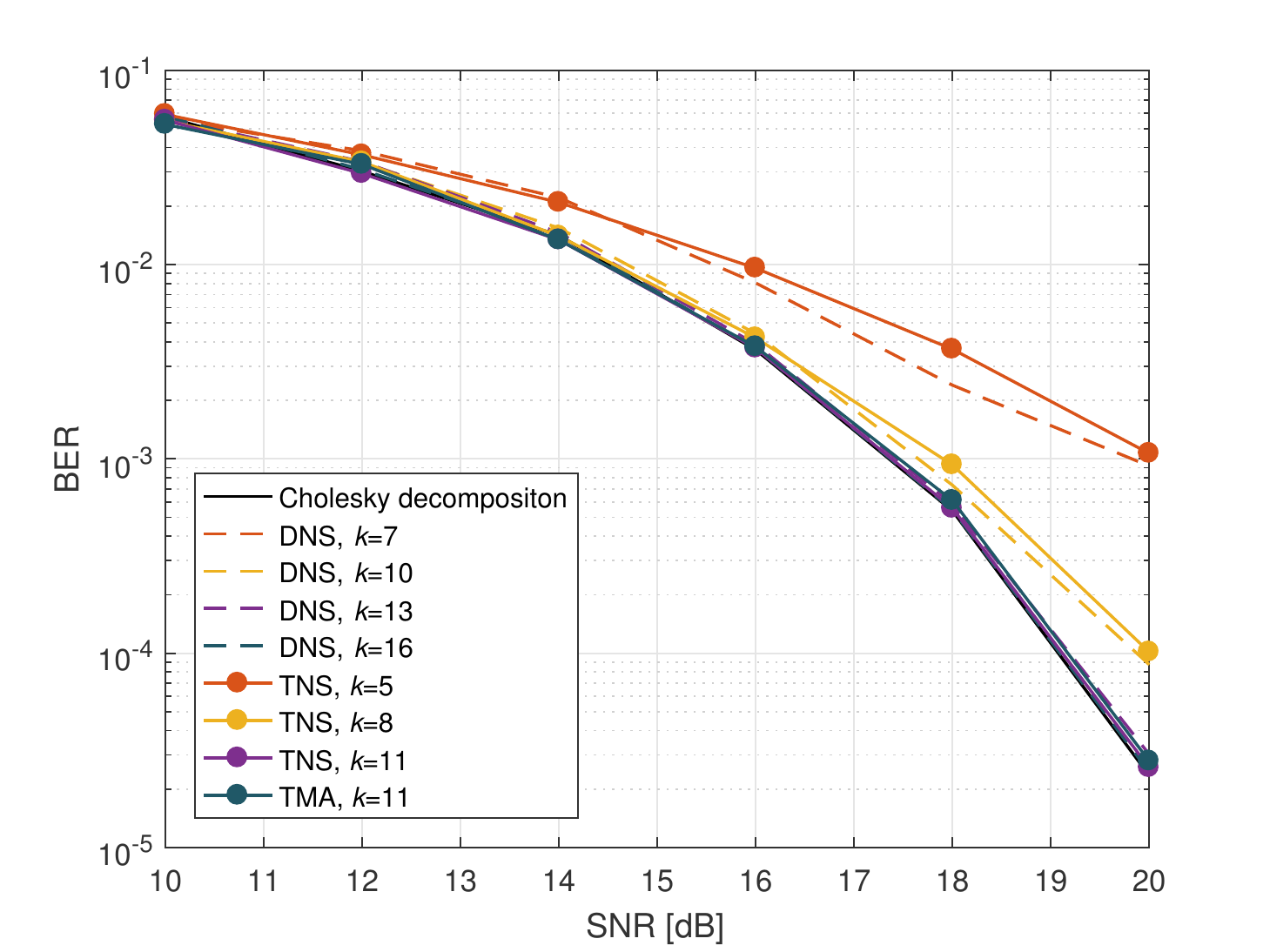}
\caption{\textcolor{black}{BER comparison among exact MMSE method, DNS, TNS, and TMA. Set $\beta=N/K=128/16=8$ in $\zeta=0.3$ channel by $64$-QAM.}}
\label{fig:r_4}
\end{figure}
\begin{figure}[ht]
\centering
\includegraphics[width=0.42\textwidth]{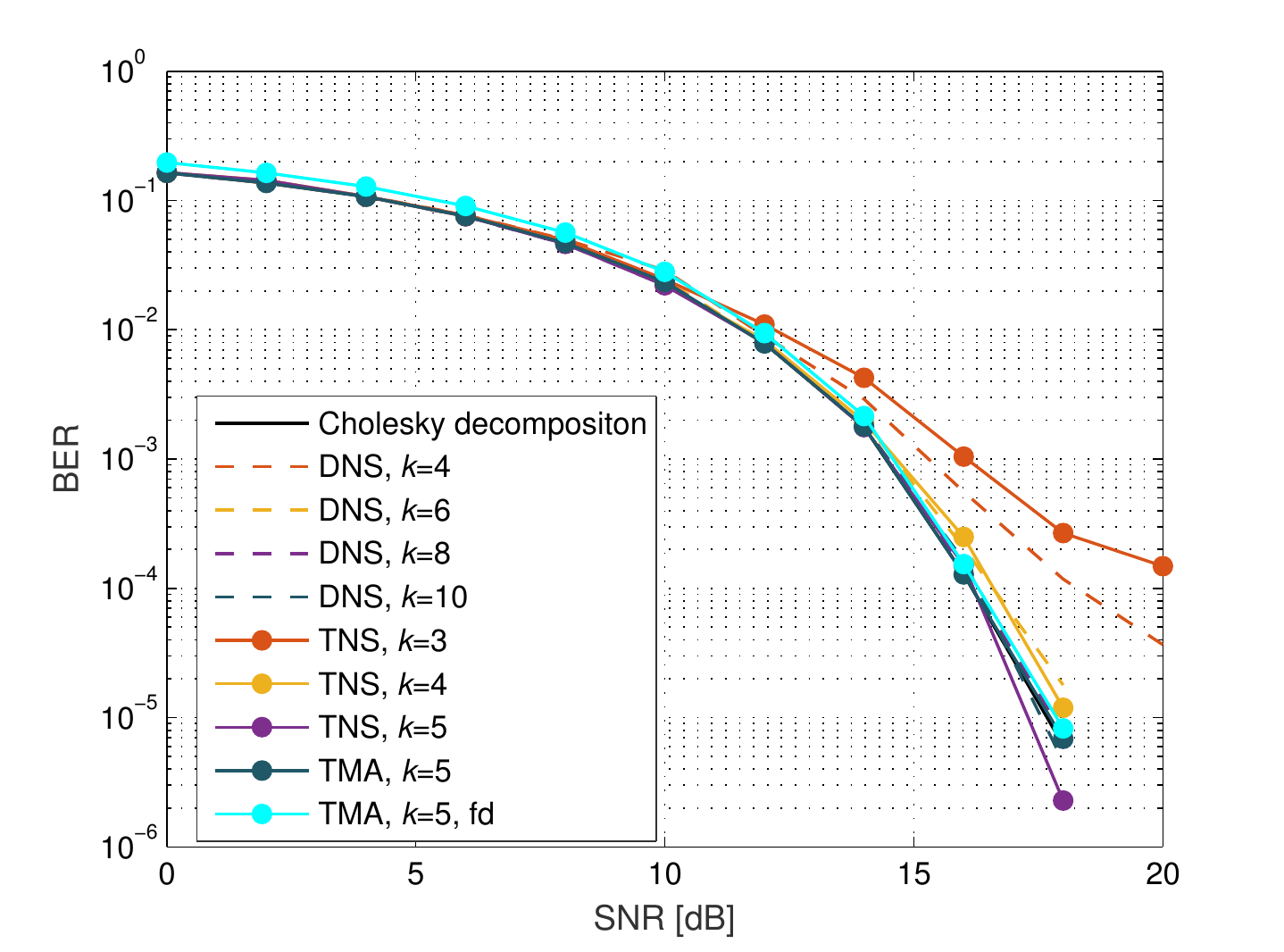}
\caption{\textcolor{black}{BER comparison among exact MMSE method, DNS, TNS, and TMA. Set $\beta=N/K=128/16=8$ in $\zeta=0.5$ channel by $64$-QAM.}}
\label{fig:r_8}
\end{figure}

\textcolor{black}{Fig. \ref{fig:r0}-\ref{fig:r_8} show the performance comparison of the aforementioned methods in different systems, and the results are summarized in Table \ref{tab:it}.} When $\zeta$ grows, both methods need more iteration counts to reach required accuracy. According to error analysis in Section \ref{sec:err}, when $\zeta\leq 0.6$ the Neumann series scheme will not be suitable for such channel situation, our simulation results also show this in Table \ref{tab:it}. Another observation is that larger $\beta=N/K$ requires fewer iteration counts. From the comparison we can clearly find that TMA saves $1$ to $3$ iterations than DNS. Also, our improved approximate tridiagonal matrix inversion in Section \ref{sec:step1} does not bring extra performance loss.

\begin{figure}[ht]
\centering
\includegraphics[width=0.42 \textwidth]{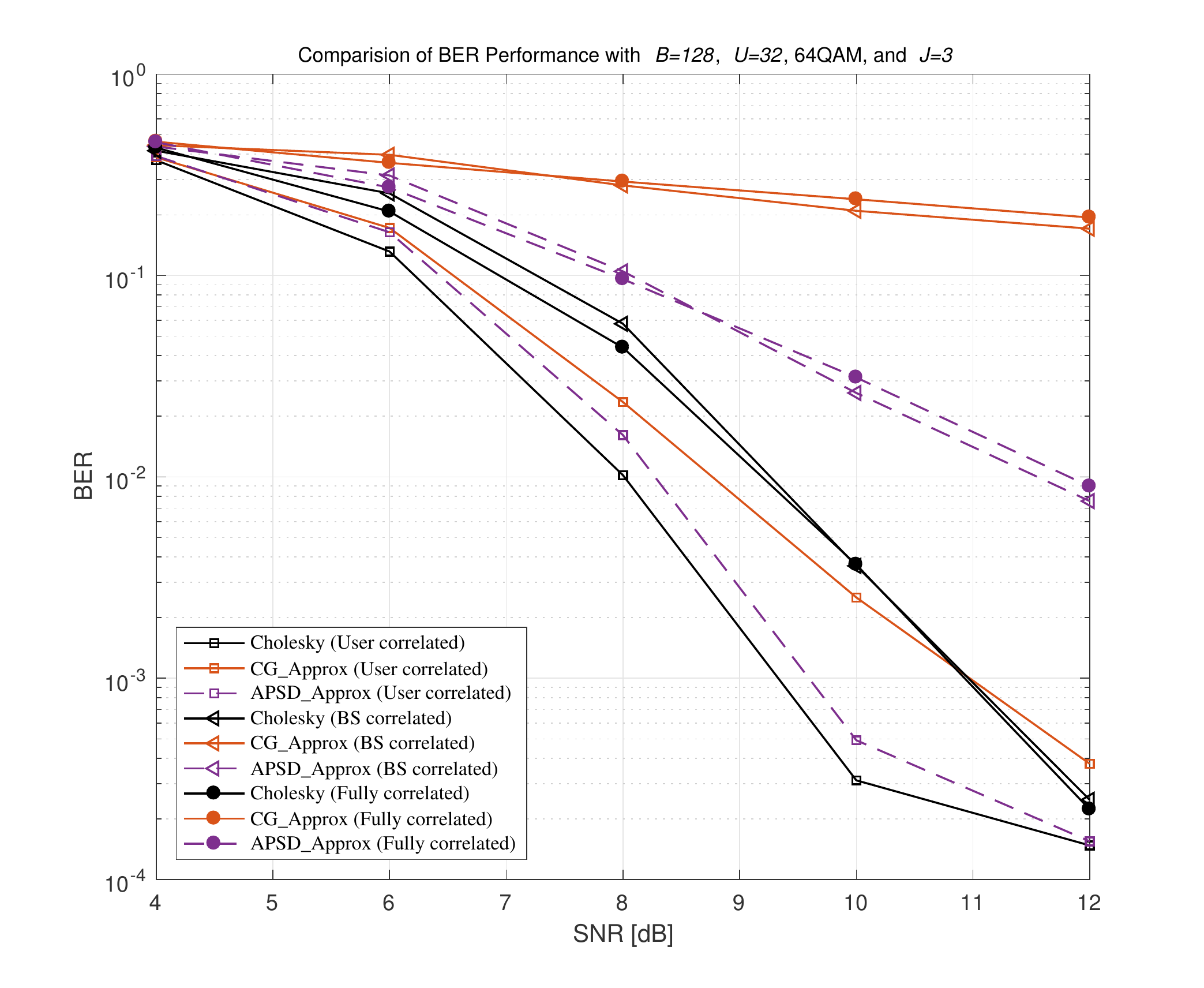}
\caption{Performance comparison with different correlation conditions.}
\label{fig:difi}
\end{figure}

\textcolor{black}{In order to better show the advantage of proposed methods, it is necessary to compare their performance with other algorithms. Rather than put 1 or 2 other algorithms in each of Fig.s \ref{fig:r0}-\ref{fig:r8}, we would like to show that under different correlation scenarios, the state-of-the-art linear iteration detectors such as CG or SD (adaptive preconditioned SD, APSD) methods could not outperform the exact MMSE detector based on Cholesky decomposition. As long as our proposed detectors can achieve comparable performance as the exact MMSE detector based on Cholesky decomposition, their performance is guaranteed even compared with the state-of-the-art linear iterative detectors. In Fig. \ref{fig:difi}, BER performances under different correlation conditions are compared. Specifically, we have: \romannumeral1) {\em User Correlated} case $(\zeta_t=0.2, \zeta_r=0)$, \romannumeral2) {\em BS Correlated} case $(\zeta_t=0, \zeta_r=0.4)$, and \romannumeral3) {\em Fully correlated} case $(\zeta_t=0.2, \zeta_r=0.4)$.
It is shown that, the proposed TMA detector can effectively approach the exact MMSE detector in different correlation conditions. Therefore, the advantages of proposed decoders can be guaranteed over the state-of-the-art detectors, regarding different correlation conditions.}

\subsection{VLSI Implementation Results}
We investigate the VLSI implementation architecture for the proposed (TMA and DNS friendly) detection method based on a Xilinx Virtex-7 XC7VX690T FPGA and compare the hardware cost in Table \ref{tab:fpgare}. For fair comparison with NSE \cite{MichaelWu:14}, GS \cite{Me2016iscas}, CG \cite{Yin2015iscas}, and OCD \cite{Wu2016cd}, the quantization length is set as $15$. \textcolor{black}{The fixed-point simulation results of proposed quantization scheme are shown in Fig.s \ref{fig:r0}-\ref{fig:r8}, and the performance loss is negligible. Here, ``fd'' is short for ``fixed''.} We assume $N=128$ BS antennas and $K=8$ users. Although TMA detection requires $134\%$ LUTs as DNS, it requires lower $L$ and enjoys higher throughput. Shown in Tables \ref{tab:comp1} and \ref{tab:comp2}, with decreased $\beta$ and increased $\zeta$, the throughput advantage is more obvious.
\begin{table}[htbp]
\tabcolsep 1mm
\renewcommand{\arraystretch}{1.2}
\centering
\footnotesize
\caption{FPGA results for TMA and DNS with $N/K=128/8$}
\label{tab:fpgare}
\begin{tabular}{lll}
\Xhline{1.0pt}
Detection algorithm & TMA & DNS\\
\hline
\rowcolor{mygray}
Modulation & $64$-QAM & $64$-QAM\\
LUTs	& $9,795$ $(2.27\%)$    & $7,261$ $(1.72\%)$ \\
\rowcolor{mygray}
FFs   & $5,874$ $(0.69\%)$   &  $4,524$ $(0.52\%)$ \\
DSP48s & $200$ $(5.56\%)$ & $144$ $(4.00\%)$\\
\rowcolor{mygray}
Block RAMs & $4$   & $4$  \\
Maximum frequency & $225$ MHz  & $225$ MHz \\
\Xhline{1.0pt}
\end{tabular}
\end{table}

\begin{table}[ht]
    \tabcolsep 1mm
    \renewcommand{\arraystretch}{1.2}
    \footnotesize
    \caption{Throughput for different $\zeta$'s, when $\beta=16$, $K=8$, $64$-QAM}
    \begin{center}
    \begin{tabular}{c || c | c | c | c | c | c }
    \Xhline{1.0pt}
    \multirow{2}{*}{Method}  &\multicolumn{3}{c|}{Latency (clocks)}&\multicolumn{3}{c}{Throughput (Mbps)}\\
    \cline{2-7}& $\zeta=0$ & $\zeta=0.3$ & $\zeta=0.5$ & $\zeta=0$ &$\zeta=0.3$ & $\zeta=0.5$\\
    \hline
     DNS &  $176$ & $185$ & $212$ & $61.36$& $58.37$ & $50.93$ \\
    \hline
     TMA  &  $169$ & $178$ & $187$ & $63.90$& $60.67$ & $57.74$ \\
    \Xhline{1.0pt}
    \end{tabular}\label{tab:comp1}
    \end{center}
    \end{table}

\begin{table}[ht]
    \tabcolsep 1mm
    \renewcommand{\arraystretch}{1.2}
    \footnotesize
    \caption{Throughput for different $\zeta$'s, when $\beta=8$, $K=16$, $64$-QAM}
    \begin{center}
    \begin{tabular}{c || c | c | c | c | c | c }
    \Xhline{1.0pt}
    \multirow{2}{*}{Method} &\multicolumn{3}{c|}{Latency (clocks)}&\multicolumn{3}{c}{Throughput (Mbps)}\\
    \cline{2-7}& $\zeta=0$ & $\zeta=0.3$ & $\zeta=0.5$ & $\zeta=0$ &$\zeta=0.3$ & $\zeta=0.5$\\
    \hline
     DNS &  $203$ & $257$ & $392$ & $53.20$ & $42.01$ & $27.54$ \\
    \hline
     TMA  &  $196$ & $241$ & $268$ & $55.10$& $44.80$ & $37.95$ \\
    \Xhline{1.0pt}
    \end{tabular}\label{tab:comp2}
    \end{center}
    \end{table}

{\color{black}{To achieve $75$ Mb/s data rate for each user specified in 3GPP LTE-Advanced \cite{lte}, we use $10$ instances of TMA detector.}} Table \ref{tbl:results} lists comparison with other detection implementation results with Xilinx Virtex-7 XC7VX690T FPGA for $128\times 8$ MIMO system: NSE \cite{MichaelWu:14}, GS \cite{Me2016iscas}, CG \cite{Yin2015iscas}, and OCD \cite{Wu2016cd}. It is shown that for correlated massive MIMO systems, it can achieve near-MMSE performance and $630$ Mb/s throughput. For fair comparison, hardware efficiency is compared in terms of throughput/LUTs and throughput/FFs. The throughput/LUTs ratio of proposed TMS is better than NSE \cite{MichaelWu:14}, GS \cite{Me2016iscas}, and CG \cite{Yin2015iscas}. If we consider another metric throughput/FF, it would be noted that this work will have the highest value of $9,027$ compared to the state-of-the-art.
\begin{table*}[tbp]
\tabcolsep 2mm
\renewcommand{\arraystretch}{1.2}
\centering
\footnotesize
\caption{Implementation results and comparison on a Xilinx Virtex-7 XC7VX690T FPGA for $128\times 8$ MIMO system}
\label{tbl:results}
\begin{tabular}{llllll}
\Xhline{1.0pt}
\multirow{2}{*}{Detector}          & M. Wu \cite{MichaelWu:14}       & Z. Wu \cite{Me2016iscas}              & B. Yin \cite{Yin2015iscas}          & M. Wu \cite{Wu2016cd}    & \multirow{2}{*}{This Work}  \\
& [JSTSP'Nov. 14]       & [ISCAS'16]              & [ISCAS'15]          & [TCAS-I'Dec. 16]    &  \\\hline
\rowcolor{mygray}
Method          & NSE       & GS              & CG          & OCD    & TMA  \\
Performance       & near-MMSE & near-MMSE  & near-MMSE & near-MMSE & near-MMSE \\
\rowcolor{mygray}
Modulation scheme & $64$-QAM    & $64$-QAM    & $64$-QAM    & $64$-QAM       & $64$-QAM     \\
Iteration count   & $3$         & $1$         & $3$         & $3$       & $3$   \\
\rowcolor{mygray}
Instance or subcarrier \#   & $8$         & $1$         & $1$         & $24$       & $10$   \\ \hline
LUTs				&$148,797$	&$18,976$			&$3,324$			&$23,914$		&$91,353$\\
\rowcolor{mygray}
FFs					&$161,934$	&$15,864$				&$3,878$				&$43,008$	&$69,784$\\
DSP48s				&$1,016$		&$232$				&$33$			&$774$		&$2,000$	\\
\rowcolor{mygray}
BRAMs				&$16$				&$6$						&$1$							&$2$		&$40$			\\ \hline
Latency [clks]		&$196$				&$311$				&$951$							&$795$		&$169$			\\
\rowcolor{mygray}
Clock freq. [MHz]	&$317$				&$309$				&$412$						&$258$			&$225$			\\
Throughput [Mb/s] 	&$621$				&$48$				&$20$								&$376$	&$630$			 \\ \hline
\rowcolor{mygray}
Throughput/LUTs     &$4,173$             &$2,530$             &$6,017$               &$15,597$     & $6,896$    \\
   \Xhline{1.0pt}
\end{tabular}
\end{table*}

\subsection{Comments and Further Work}
\emph{Advantages}: The hardware cost of DNS is lower than that of accurate inversion method (Cholesky decomposition). With weaker requirement of iteration number, DNS enjoys a higher throughput. With the same performance, TMA saves $2$ iterations than DNS. Since the iteration number can be easily controlled, it is feasible for different applications.

\emph{Drawbacks}: On bad channel condition, both DNS and TMA detections need higher iteration counts $L$, which increases the detecting latency and lowers the throughput.

\emph {An improved Neumann series}: In order to solve the problem, an improved Neumann series inverse framework is proposed for higher convergence rate. The formula of Neumann series can be rewritten as follows:
\begin{equation}
\sum_{n=0}^{2^{L\!-\!1}}\!{\mathbf{\Theta}^n\mathbf{X}^{-1}}\!=\!\mathbf{\Theta}^{2^{L\!-\!2}}\!(\sum_{n=0}^{2^{L\!-\!2}-1}\!{\mathbf{\Theta}^n\mathbf{X}^{-1}})
\!+\!(\sum_{n=0}^{2^{L\!-\!2}-1}\!{\mathbf{\Theta}^n\mathbf{X}^{-1}})
\end{equation}

\begin{equation}
\mathbf{\Theta}=(\mathbf{I}-\mathbf{X}^{-1}\mathbf{W})
\end{equation}
Then a formula based on multiple iterations is obtained:
\begin{equation}
\mathbf{W}_L^{-1}=\left\{
             \begin{aligned}
              &\mathbf{X}^{-1},~L=1 \\
              &(\mathbf{I}-\mathbf{X}^{-1}\mathbf{W})^{2^{L-2}}\mathbf{W}_{L-1}^{-1}+\mathbf{W}_{L-1}^{-1},~L \geq 2\\
             \end{aligned}
           \right.
\end{equation}

\begin{figure}[ht]
\centering
\includegraphics[width=0.42 \textwidth]{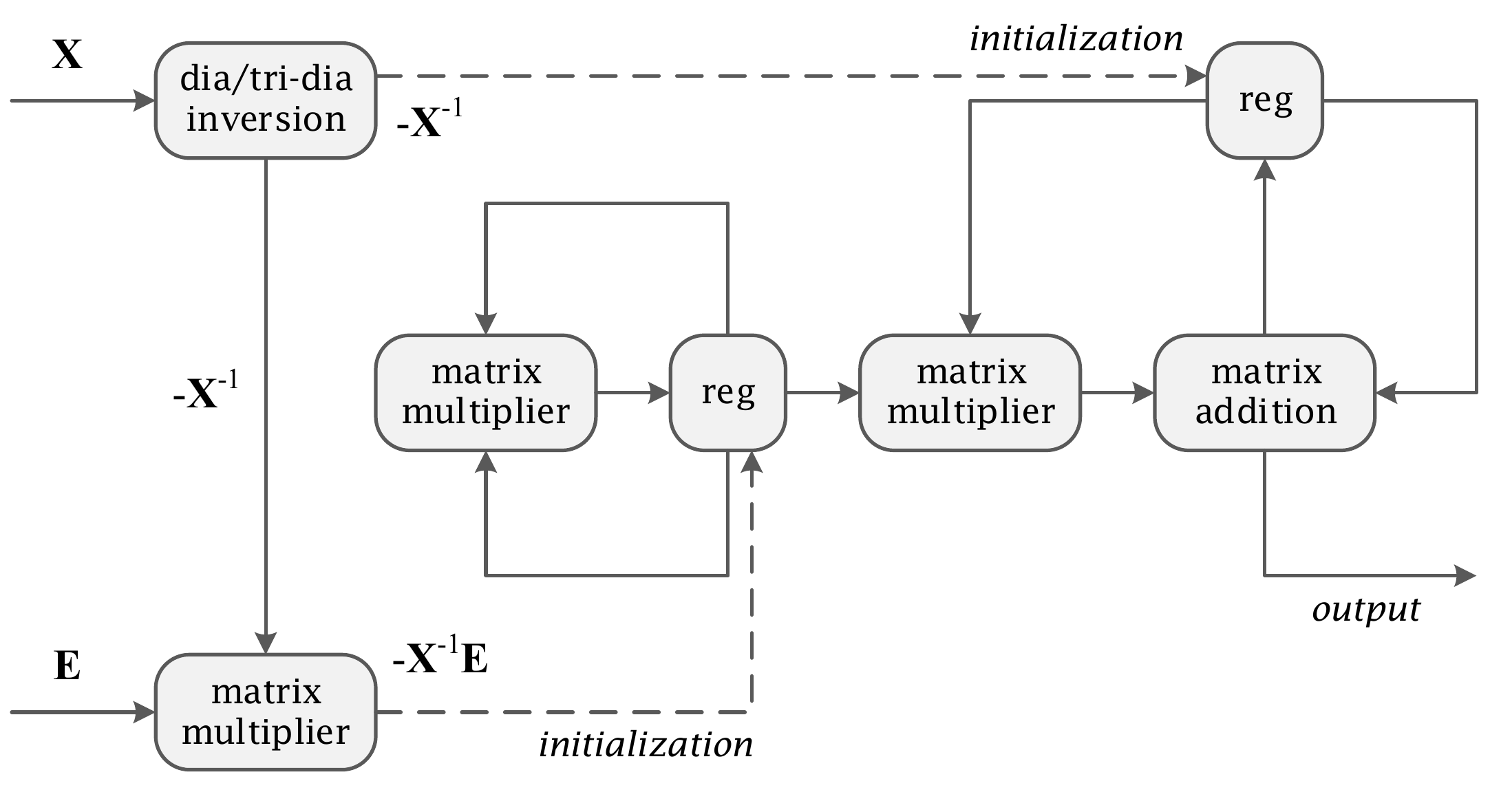}
\caption{Architecture of the improved approximate matrix inversion.}
\label{fig:ARC2}
\end{figure}

To obtain the $L^{\text{th}}$ iteration's output of original Neumann series, now we only need $2^{L-1}$ iterations. Convergence comparison after such alteration is listed in Table \ref{tab:COMPARE}. Compared with the original iteration inverse framework, the modified scheme adds a systolic multiplication module. Though it requires more hardware, a great rise of convergence and throughput appears. Also, the hardware can be parallelized for lower latency per iteration. This framework applies to situations with harsh channel condition and relatively high accuracy requirement. The iteration part of modified hardware framework is in Fig. \ref{fig:ARC2}.

There are mainly two differences from the former structure. Firstly, the former constant multiplication coefficient matrix will be replaced by continuous self-squares. Secondly, the bias increment matrix, which used to be fixed in the original iteration, is now replaced by the last result of iteration operation.
 \begin{table*}[ht]
\tabcolsep 2mm
\renewcommand{\arraystretch}{1.3}
\centering
\footnotesize
\caption{Convergence comparison between original and improved Neumann series}
\begin{tabular}{l | l l | l l}
\Xhline{1.0pt}
Iteration & Original Neumann series & Order & Improved Neumann series & Order  \\
\hline
\rowcolor{mygray}
$1$ & $\mathbf{W}^{-1}$ & $1$ & $\mathbf{W}^{-1}$ & $1$  \\
$2$ & $\mathbf{\Theta}\mathbf{W}^{-1}+\mathbf{W}^{-1}$ & $2$ &
$\mathbf{\Theta}\mathbf{W}^{-1}+\mathbf{W}^{-1}$ & $2$  \\
\rowcolor{mygray}
$3$ & $\mathbf{\Theta}^2\mathbf{W}^{-1}+\mathbf{\Theta}\mathbf{W}^{-1}+\mathbf{W}^{-1}$ & $3$ & $\mathbf{\Theta}(\mathbf{\Theta}\mathbf{W}^{-1}+\mathbf{W}^{-1})+(\mathbf{\Theta}\mathbf{W}^{-1}+\mathbf{W}^{-1})$ & $4$ \\
$4$ & $\sum_{n=0}^3{\mathbf{\Theta}^n\mathbf{W}^{-1}}$ & $4$ & $\mathbf{\Theta}^4(\sum_{n=0}^3{\mathbf{\Theta}^n\mathbf{W}^{-1}})+(\sum_{n=0}^3{\mathbf{\Theta}^n\mathbf{W}^{-1}})$ & $8$ \\
\rowcolor{mygray}
$L$ & $\sum_{n=0}^{L-1}{\mathbf{\Theta}^n\mathbf{W}^{-1}}$ & $L$ & $\mathbf{\Theta}^{2^{L-2}}(\sum_{n=0}^{2^{L-2}-1}{\mathbf{\Theta}^n\mathbf{W}^{-1}})+(\sum_{n=0}^{2^{L-2}-1}{\mathbf{\Theta}^n\mathbf{W}^{-1}})$ & $2^{L-1}$ \\
\Xhline{1.0pt}
\end{tabular}\label{tab:COMPARE}
\end{table*}

\section{Conclusion}
This paper proposes a modified NSE based on tridiagonal matrix inversion approximation called TMA method. It well accommodates the complexity as well as the performance issues for correlated M-MIMO systems. Its approximation errors, computational complexity, and processing latency have been analyzed. Meanwhile, we investigate the VLSI architecture for the proposed algorithm and implemented it on a Xilinx Virtex-7 XC7VX690T FPGA. Compared with benchmark systems, our proposed efficient pipelined TMA detector can get high hardware efficiency. Finally, we also propose a fast iteration structure for further research.


\appendices
\section{Proof of Theorem \ref{thm:cor}} \label{pf:thm:cor}
Let $\mathbf{H}\in \mathbb{C}^{N \times K}$ be i.i.d. \textit{circularly symmetric complex Gaussian} with zero mean and unit variance, where
\begin{equation}
  \left\{
  \begin{aligned}
    &\mathbf{H}=[\widetilde{\mathbf{H}}_{1},\widetilde{\mathbf{H}}_{2},\cdots,\widetilde{\mathbf{H}}_{K}]_{N\times K},\\
    &\widetilde{\mathbf{H}}^{H}_{i}=[h_{1i}^{*},h_{2i}^{*},\cdots,h^{*}_{Ni}]_{1\times N}.
      \end{aligned}
      \right.
\end{equation}

For $\mathbf{W}=(\mathbf{R}^{1/2}\mathbf{H}\mathbf{\Sigma}^{1/2})^{H}
(\mathbf{R}^{1/2}\mathbf{H}\mathbf{\Sigma}^{1/2})+\eta^{2}\mathbf{I}_{K}$, the filtering matrix, its elements are expressed as
\begin{equation}
\left\{
\begin{aligned}
w_{ii}&=\mathbf{\sigma}_{i}\widetilde{\mathbf{H}}_{i}^{H}\mathbf{R}\widetilde{\mathbf{H}}_{i}+\eta^{2},~i=1,\ldots,K  \\
w_{ij}&=\sqrt{\sigma_{i}\sigma_{j}}\widetilde{\mathbf{H}}_{i}^{H}\mathbf{R}\widetilde{\mathbf{H}}_{j},~i,j=1,\ldots,K,j\neq i.
\end{aligned}
\right.
\label{equ:ele}
\end{equation}

Given the denotation of $\mathbf{R}=\{(\zeta e^{j\omega_{lm}})^{|l-m|}\}$, $\widetilde{\mathbf{H}}^{H}_{i}\mathbf{R}\widetilde{\mathbf{H}}_{j}$ is further presented as
\begin{equation}\label{eqn:hRh}
\widetilde{\mathbf{H}}^{H}_{i}\mathbf{R}\widetilde{\mathbf{H}}_{j}
=\sum_{m=1}^{N}{\sum_{l=1}^{N}{h^{*}_{li}h_{mj}(\zeta e^{j\omega_{lm}})^{|l-m|}}}.
\end{equation}

\subsection{Proof of the Expectation of $\mathbf{W}$}
The mathematical expectation of $w_{ii}$ is $\mathbb{E}\{w_{ii}\}$. When $N$ grows, only products of correlated terms remain significant. Therefore,
\begin{equation}\label{eqn:ewii}
\resizebox{.88\hsize}{!}{$
\begin{aligned}
\mathbb{E}\{w_{ii}\}
&=\sigma_{i}\mathbb{E}\{\sum_{l=1}^{N}{|h_{li}|^2}\}\\
&+\mathbb{E}\{\sum_{m=1}^{N}{\sum_{l=1}^{N}{h^{*}_{li}h_{mi}(\zeta e^{j\omega_{lm}})^{|l-m|}}}\}(l\neq m)+ \eta^2 \\
&=\sigma_{i}N+\eta^2\approx \sigma_{i}N.
\end{aligned}
$}
\end{equation}

Similarly, the mathematical expectation of $w_{ij}$ $(i\neq j)$ is
\begin{equation}\label{eqn:ewij}
\resizebox{.89\hsize}{!}{$
\mathbb{E}\{w_{ij}\} =\sqrt{\sigma_{i}\sigma_{j}}\mathbb{E}\{\sum_{m=1}^{N}{\sum_{l=1}^{N}{h^{H}_{li}h_{mj}(\zeta e^{j\omega_{lm}})^{|l-m|}}}\} =0.
$}
\end{equation}

Again, when $N$ grows large, substituting Eq.s (\ref{eqn:ewii}) and (\ref{eqn:ewij}), we obtain the mean, $\mathbb{E}(\mathbf{W}/N)=\mathbf{\Sigma}$.\qed

\subsection{Proof of the Variance of $\mathbf{W}$}
For the variance of $w_{ij}$ $(i\neq j)$,
\begin{equation}
\begin{aligned}
\mathbf{\Omega}_{w_{ij}}/\sigma_{i}\sigma_{j} & = \mathbb{E}\{w_{ij}^{2}\}/\sigma_{i}\sigma_{j}-\mathbb{E}\{w_{ij}\}^2/\sigma_{i}\sigma_{j}  \\
&= \mathbb{E}\{ (\widetilde{\mathbf{H}}_{i}^{H}\mathbf{R}\widetilde{\mathbf{H}}_{j})
(\widetilde{\mathbf{H}}_{i}^{H}\mathbf{R}\widetilde{\mathbf{H}}_{j})^{H}\}.
\end{aligned}
\end{equation}

Substitute Eq. (\ref{eqn:hRh}), only the correlated terms remain significant, and eliminate the zero terms, we get
\begin{equation}
\begin{aligned}
&\mathbb{E}\{(\widetilde{\mathbf{H}}_{i}^{H}\mathbf{R}\widetilde{\mathbf{H}}_{j})
(\widetilde{\mathbf{H}}_{i}^{H}\mathbf{R}\widetilde{\mathbf{H}}_{j})^{H}\}\\
=&\mathbb{E}\{\sum_{m=1}^{N}{\sum_{l=1}^{N}{|h_{li}|^{2}|h_{mj}|^{2}\zeta ^{2|l-m|}}}\}.
\end{aligned}
\end{equation}
Then, $\mathbf{\Omega}_{w_{ij}}/\sigma_{i}\sigma_{j} =\sum_{m=1}^{N}{\sum_{l=1}^{N}{\zeta^{2|l-m|}}}= \|\mathbf{R}(\zeta)\|^2_{F}$ is the variance of  $w_{ij}$ $(i\neq j)$, where $\|\mathbf{R}(\zeta)\|^2_{F}$ is an increasing function of $\zeta$ with the minimum value, $\min(\|\mathbf{R}(\zeta)\|^2_{F})=0$ when $\zeta=0$, and the maximum value, $\max(\|\mathbf{R}(\zeta)\|^2_{F})=N^2$ when $\zeta=1$.

Further, as $N$ grows large, the effect of uncorrelated noise is averaged out. We obtain the variance of $w_{ii}$ as:
\begin{equation}
\begin{aligned}
\mathbf{\Omega}_{w_{ii}}/\sigma_{i}^2
 &= \mathbb{E}\{w_{ii}^{2}\}/\sigma_{i}^2-\mathbb{E}\{w_{ii}\}^2/\sigma_{i}^2  \\
 &\approx \mathbb{E}\{ (\widetilde{\mathbf{H}}_{i}^{H}\mathbf{R}\widetilde{\mathbf{H}}_{i})
(\widetilde{\mathbf{H}}_{i}^{H}\mathbf{R}\widetilde{\mathbf{H}}_{i})^{H}\} -N^2.
\end{aligned}
\end{equation}

We expand the inner polynomial and eliminate the zero terms as follows
\begin{equation}
\begin{aligned}
&\mathbb{E}\{ (\widetilde{\mathbf{H}}_{i}^{H}\mathbf{R}\widetilde{\mathbf{H}}_{i})
(\widetilde{\mathbf{H}}_{i}^{H}\mathbf{R}\widetilde{\mathbf{H}}_{i})^{H}\}\\
 =&\mathbb{E}\{\sum_{l=1}^{N}{|h_{li}|^{4}}\}+\mathbb{E}\{\sum_{l=1}^{N,l \neq m}{{|h_{li}|^{2}}
\sum_{m=1}^{N}{|h_{mi}|^{2}}}\}+ \\ & \mathbb{E}\{\sum_{l=1}^{N,l \neq m}{\sum_{m=1}^{N}|h_{li}|^2|h_{mi}|^2\zeta ^{2|l-m|}}\}.
\end{aligned}
\end{equation}

Note that the fact $\mathbb{E}\{|h_{ij}|^{4}\}=2$, we obtain
\begin{equation}
\resizebox{.89\hsize}{!}{$
\begin{aligned}
\mathbf{\Omega}_{w_{ii}}/\sigma_{i}^2 &= 2N+N(N-1)+ \sum_{l=1}^{N,l\neq m}{\sum_{m=1}^{N}{\zeta ^{2|l-m|}}}-N^2\\
&= N + \sum_{l=1}^{N,l\neq m}{\sum_{m=1}^{N}{\zeta ^{2|l-m|}}}
= \|\mathbf{R}(\zeta)\|^2_{F}.
\end{aligned}
$}
\end{equation}

Finally, we acquire the variance of $\mathbf{W}$ as
\begin{equation}
\begin{aligned}
\mathbf{\Omega}_{\mathbf{W}}
& = \mathbb{E} \left[ \left(\mathbf{W - \mathbf{\Sigma}_{\mathbf{W}}} \right) \left(\mathbf{W - \mathbf{\Sigma}_{\mathbf{W}}}\right)^{H}\right]\\
& = \|\mathbf{R}(\zeta)\|^2_{F}
\mathbf{\Sigma}\overline{\mathbf{I}}_{K}\mathbf{\Sigma}.
\end{aligned}
\end{equation}

We can see the angle $\omega$ does not affect system performance. Therefore, system performance is dominated by the magnitude of correlation, $\zeta$.\qed

\section{Proof of $\mathbb{E}\{\Phi\}$} \label{pf:thm:phi}
Consider the truncated NSE, $\mathbf{W}^{-1}(L)=\sum_{n=0}^{L-1}{\mathbf{\Theta}^{n}\widehat{\mathbf{W}}}$, we get $\mathbf{\Delta}_{\mathbf{W}|L}=\mathbf{W}^{-1}-\mathbf{W}_{L}^{-1}=\mathbf{\Theta}^{L}\mathbf{W}^{-1}$.

Substitute $\mathbf{\Delta}_{\mathbf{W}|L}=\mathbf{\Theta}^{L}\mathbf{W}^{-1}$ and Eq. (\ref{eqn:est}) into $\Phi$. Then, we acquire $\Phi =\left\|\mathbf{\Theta}^{L}\hat{\mathbf{s}}\right\|_{2}^{2}$, where $\hat{\mathbf{s}}=$ is the exact estimate. Here, the $\Phi$ is upper bounded as $\Phi\leq \|\mathbf{\Theta}\|_{F}^{2}\|\hat{\mathbf{s}}\|_{2}^{2}$.

In the typical propagation conditions for M-MIMO systems, $N$ grows and $\beta$ is large, the effect of uncorrelated noise could be averaged out. In this case, applying the denotation of $2$-norm, $\mathbb{E}\{\Phi\}$ can be expressed as
\begin{equation}\label{equ:Eest}
\begin{aligned}
\mathbb{E}\{\Phi\}&=\mathbb{E}(\left\|\mathbf{\Theta}^{L}\mathbf{s}\right\|^{2}_{2})=\mathbb{E}\{\mathrm{Tr}[\mathbf{\Theta}^{L}\mathbf{s}
(\mathbf{\Theta}^{L}\mathbf{s})^{H}]\}\\
    &=\mathbb{E}\{\mathrm{Tr}[\mathbf{s}\mathbf{s}^{H}(\mathbf{\Theta}^{L})^{H}\mathbf{\Theta}^{L}]\}.
\end{aligned}
\end{equation}
Note that $\mathbb{E}\{|s_{k}^{2}|\}=1$, and $\mathbf{s}$ is independent to $\mathbf{\Theta}$. Eq. (\ref{equ:Eest}) can be derived as
\begin{equation}
\begin{aligned}
\mathbb{E}\{\Phi\}
    &=\mathrm{Tr}\{\mathbb{E}(\mathbf{s}\mathbf{s}^{H})\mathbb{E}[(\mathbf{\Theta}^{L})^{H}\mathbf{\Theta}^{L}]\}\\
    &=\mathrm{Tr}\{\mathbf{I}_{K}\mathbb{E}[(\mathbf{\Theta}^{L})^{H}\mathbf{\Theta}^{L}]\}\\
    &=\mathbb{E}\{\left\|\mathbf{\Theta}^{L}\right\|^{2}_{F}\}.\qed
\end{aligned}
\end{equation}

\section{Proof of $\mathbf{W}^{-1}(L)$ as a Hermitian matrix } \label{pf:thm:her}
The expanded formula of Neumann Series Approximation Inverse Eq. (\ref{equ:iter}) can be written as the following form,
\begin{equation}
\resizebox{.89\hsize}{!}{$
\begin{aligned}
\mathbf{W}^{-1}(L) =& (-1)^{L-1}\underbrace{\mathbf{X}^{-1}\mathbf{E}\dots\mathbf{X}^{-1}\mathbf{E}\mathbf{X}^{-1}\dots\mathbf{E}\mathbf{X}^{-1}}_{2L-1}\\
                    &+(-1)^{L-2}\underbrace{\mathbf{X}^{-1}\mathbf{E}\dots\mathbf{X}^{-1}\mathbf{E}\mathbf{X}^{-1}\dots\mathbf{E}\mathbf{X}^{-1}}_{2L-3}\\
                    &\dots\\
                    &-\mathbf{X}^{-1}\mathbf{E}\mathbf{X}^{-1}+\mathbf{X}^{-1},
\end{aligned}
$}
\end{equation}
where $\mathbf{X}^{-1}=\mathbf{W}^{-1}_{\textrm{dia/tri}}$. For $\mathbf{X}^{-1}$ and $\mathbf{E}$ are Hermitian, then
\begin{equation}
\resizebox{.89\hsize}{!}{$
\begin{aligned}
&(\mathbf{X}^{-1}\mathbf{E}\dots\mathbf{X}^{-1}\mathbf{E}\mathbf{X}^{-1}\dots\mathbf{E}\mathbf{X}^{-1})^{H}\\
=&(\mathbf{X}^{-1})^{H}(\mathbf{E})^{H}\dots(\mathbf{X}^{-1})^{H}(\mathbf{E})^{H}(\mathbf{X}^{-1})^{H}\dots(\mathbf{E})^{H}(\mathbf{X}^{-1})^{H}\\
=&\mathbf{X}^{-1}\mathbf{E}\dots\mathbf{X}^{-1}\mathbf{E}\mathbf{X}^{-1}\dots\mathbf{E}\mathbf{X}^{-1}.
\end{aligned}
$}
\end{equation}
Finally, we can proof that $\mathbf{W}^{-1}(L)$ is a Hermitian matrix, $(\mathbf{W}^{-1}(L))^{H}=\mathbf{W}^{-1}(L)$.\qed
\section{\textcolor{black}{Derivation of Algorithm \ref{alg:triapp}} } \label{pf:thm:all}
After two steps of simplification in Sections \ref{sec:step1} and \ref{sec:step2}, the formula is written as follows,
\begin{equation}\label{eqn:znew}
z_i=b_iz_{i-1}-|a_i|^2z_{i-2},~i=2,3,\ldots,K,
\end{equation}
where $z_0=1$, $z_1=b_1$. Also

\begin{equation}\label{eqn:phnew:1}
\phi_{ij}=
\left\{
\begin{aligned}
&\frac{1}{b_i-|a_i|^2\frac{z_{i-2}}{z_{i-1}}-\frac{|a_{i+1}|^2}{b_{i+1}}},~ {i = j}, \\
&-a_i^*\frac{z_{j-2}}{z_{j-1}}\phi_{jj},~ {i = j-1}, \\
&\phi_{ji}^*,~ {i = j+1}, \\
&0,~ \mathrm{others}.
\end{aligned}
\right.
\end{equation}

The formula can be deduced further. Set $q_i=b_i-|a_i|^2\frac{z_{i-2}}{z_{i-1}}$. According to Eq. (\ref{eqn:znew}),
\begin{equation}
\begin{aligned}
q_i&=b_i-|a_i|^2\frac{z_{i-2}}{z_{i-1}}\\
&=\frac{b_iz_{i-1}-|a_i|^2z_{i-2}}{z_{i-1}}\\
&=\frac{z_{i}}{z_{i-1}}.
\end{aligned}
\end{equation}
That is to say, $\frac{z_{i-2}}{z_{i-1}}=\frac{1}{q_{j-1}}$. Therefore, $q_i$ can be rewritted by recurrence as follows,
\begin{equation}
q_i=\left\{
             \begin{array}{ll}
               b_i-\frac{|a_i|^2}{q_{i-1}}, ~i=2,3,\ldots,K, \\
               b_1, ~i=1.
             \end{array}
           \right.
\end{equation}
Set $p_i$ as
\begin{equation}
p_i=\left\{
             \begin{array}{ll}
               q_i-\frac{|a_{i+1}|^2}{b_{i+1}}, ~i=1,2,\ldots,K-1, \\
               q_i, ~i=K.
             \end{array}
           \right.
\end{equation}
In all, the equation can be expressed as:
\begin{equation}\label{eqn:phnew:2}
\phi_{ij}=
\left\{
\begin{aligned}
&\frac{1}{p_i},~ {i = j}, \\
&-a_i^*\frac{1}{q_{j-1}p_j},~ {i = j-1}, \\
&\phi_{ji}^*,~ {i = j+1}, \\
&0,~ \mathrm{others}.
\end{aligned}
\right. \qed
\end{equation}
\footnotesize
\bibliographystyle{IEEEtran}
\bibliography{IEEEabrv,mybib}
\end{document}